\documentclass{article}
\usepackage{graphicx} 
\usepackage[authoryear,round]{natbib}   
\usepackage{xurl}   
\usepackage{hyperref}     
\title{Cherenkov light as a mechanism for light flashes seen by astronauts in space}

\author{
D. Švecová\textsuperscript{1} ,
P. Bobík\textsuperscript{2} ,
B. Pastirčák\textsuperscript{2}
}

\date{August 2026}

\begin{document}

\maketitle
\begin{center}
\small
\textsuperscript{1} Institute of Physics, Faculty of Science, University of P. J. Šafárik in Košice,\\
Park Angelinum 9, 040 81 Košice, Slovakia \\[4pt]
\textsuperscript{2} Institute of Experimental Physics, Slovak Academy of Sciences,\\
Watsonova 47, 040 01 Košice, Slovakia \\[4pt]
\textsuperscript{*} Corresponding authors: nikasvec@gmail.com, bobik@saske.sk
\end{center}

\begin{abstract}
Astronauts first reported light flashes during the Apollo 11 mission. These events have since been associated with cosmic rays, although their underlying mechanism remains uncertain. This study investigates possible formation processes using Geant4 simulations of cosmic ray interactions in a simplified model of the human eye.
We apply published models of human visual perception to the simulated particle induced photons and compare the resulting light flash rates with experimental observations. 
Our results indicate that, in interplanetary space and in low Earth orbit outside the South Atlantic Anomaly, Cherenkov radiation generated in the eye produces retinal responses consistent with the observed frequency of astronaut light flashes.
The dominant contributing primaries are found to be high Z nuclei, primarily iron, with additional contributions from oxygen and carbon nuclei. In contrast, conditions in the South Atlantic Anomaly suggest that additional mechanisms may be required to explain observed light flashes.
We also find that the absence of light flashes at Earth’s surface is consistent with insufficient retinal stimulation from muon induced Cherenkov radiation under typical geometrical and flux conditions.
\end{abstract}

\section{Introduction}\label{Introduction}

The first mention of the phenomenon of light flashes (LFs) was their prediction by Tobias in the early 50s \cite{Ib8, Ib9}. However, the observation of this phenomenon had to wait for almost 20 years.  During the Apollo 11 mission, the astronauts mentioned seeing a light flash after their eyes adapted to the darkness in the spacecraft. Subsequent missions reported a similar phenomenon; therefore, Apollo 14 and other missions had scheduled observations of the LFs. These observations resulted in discerning three distinct types of LFs (percentage sighting ratios in parentheses) \cite{Ib1}:

\begin{itemize}
    \item Spot or starlike, also called supernova (~66\%)
    \item Streak (~25\%)
    \item Cloud (~8\%)
\end{itemize}

For further analysis, Apollo 16 and 17 had on board the Apollo Light Flashes Moving Emulsion Detector (ALFMED) experiment \cite{Ib1}. This experiment discovered that the LFs are caused by the high-energy cosmic rays (CRs) traveling through space. However, the exact mechanism was still unknown. Thus, other space missions conducted experiments researching the mechanism of the formation of the LFs. 

An example of these experiments is the Sileye (Silicon Eye) experiment. It had three generations. Sileye-1 was on board Space Station MIR from 1995 to 1997 \cite{Ib2}, and Sileye-2 was on board MIR from 1998 to 2000 \cite{Ib2}. The results of Sileye-1 showed that the origin of the LFs is not just the protons from the CRs \cite{Ib3}. Sileye-2 observed 8 particle tracks correlated with the LFs, all with proton number greater than 3 \cite{Ib4}. The third generation of Sileye was Sileye-3, or Sileye-Alteino, on board the International Space Station ISS from 2002 until it was repurposed in the ALTEA program. This experiment combined the Silleye detectors with an electroencephalograph (EEG) \cite{Ib2}. Table \ref{tab.1} shows the number of LFs per minute seen during various experiments.

\begin{table}
 \caption{The number of LFs per minute for different space missions}
 \label{tab.1}
 \centering
    \begin{tabular}{|c|c|}
        \hline
        Mission & Number of LF/min  \\
        \hline
        Apollo 14 & 0.308$^{+}$ \\
        \hline
        Apollo 15 & 0.224$^{+}$ \\
        \hline
        Apollo 16 & 0.327$^{+}$ \\
        \hline
        Apollo 17 & 0.233$^{+}$ \\
        \hline
        Skylab (1. Obs.$^{\ast}$) & 0.343$^{\bullet}$ \\ 
        \hline
        Skylab (2. Obs.$^{\ast}$) & 2.618$^{\bullet}$ \\
        \hline
        Sileye 1 & 0.182$^{\diamond}$ \\
        \hline
        Sileye 2 & 0.135$^{\diamond}$ \\
        \hline
        Sileye 3 - Alteino & 0.095$^{\dagger}$ \\
        \hline
        ALTEA program & 0.048$^{ \triangleright}$ \\
        \hline
        \multicolumn{2}{l}{$^{\ast}$Obs.: observation; $^{+}$ \cite{Ib1};} \\
        \multicolumn{2}{l}{$^{\bullet}$ \cite{Ib9}; $^{\diamond}$ \cite{Ib5};} \\
        \multicolumn{2}{l}{$^{\dagger}$ \cite{Ib12}; $^{ \triangleright}$ \cite{Ib13};} \\
    \end{tabular}
\end{table}

These results suggest two complementary mechanisms of formation \cite{Ib5}: 

\begin{enumerate}
    \item Ionization or excitation caused by direct interaction of heavy nuclei with the retina,
    \item Knock-on particles created during proton-induced nuclear interactions in the eye.
\end{enumerate}

According to \cite{Ib10, Ib14}, the number of photons adequate to signal the brain that there is a light stimulus is around 50--150. However, from this number, only around 10 \% actually reaches the retina (approximately 5--14 photons). \cite{Ib15} states even smaller interval of 5--9 photons. The majority of the photons are lost to corneal reflection, absorption by ocular media, or pass beyond the retina \cite{Ib14}. These numbers pertain to a dark-adapted eye, which corresponds to the light flashes experiments in dark conditions; therefore, it is an ideal comparison for our simulations.

We use simulations of high-energy particle interactions with a simple model of the human eye to analyse the LFs phenomenon. We employ a widely used simulation program, Geant4 \cite{Ib6} and a few simple models to simulate this phenomenon. We will discuss the details of this model in the next section.

\section{Models}
Simulating either of the potential formation mechanisms is possible; however, the first mechanism is significantly more complicated due to the complicated design of the target. The target in the first mechanism includes the retina, the optic nerve, and other circulatory systems (such as blood vessels and veins). The target in the second mechanism is much easier to simulate due to the focus on the interactions with the water-like fluid inside the human eye. In both cases, we inject particles of various energies. The first mechanism focuses on the interaction with heavy nuclei, whereas the second mechanism concentrates more on the protons. 

Our focus was first on the second mechanism and simulations of protons interacting with the human eye. We selected a simulation software package to simulate these events. We chose Geant4 due to its comprehensive simulation toolkit, suitable for particle interaction.

Geant4 is free, object-oriented simulation software containing all the tools needed to simulate the interactions of particles with matter. In 1993, CERN and KEK independently studied the possibility of improving the previous FORTRAN-based Geant3 \cite{Mb1}. In 1994, both organizations merged their studies into one formal proposal to construct a new object-oriented simulation toolkit \cite{Mb1}. It is primarily used to simulate the progress of particles through matter.  

The following sections describe the model of the human eye created with Geant4 and all the specifications.

\subsection{Geant4 model}
As mentioned above, we used Geant4 to simulate a simple model of the human eye. Geant4 contains examples that show how to handle the toolkit and how the code of each Geant4 simulation should look. Our model was based on example B1 from Geant4's basic examples with optical parameters taken from example OpNovice from extended optical examples. We use root alongside CSV format to save and analyse the results of the simulations. All parts of the simulation code are publicly available on this site \cite{ORs1}.

The materials, chosen for our simplest simulations are air and water. We used water for the human eye because there is no predefined human eye tissue (aqueous humour, vitreous humour, iris, etc.) among Geant4's predefined materials, except for the lens \cite{Mo4}. We plan to expand the model to include these materials, similar to \cite{Mb3}, which describes a two-step approach to the definition of a material. The first step is to define all the chemicals inside the human eye. The second step is to use the AddMaterial method to create the final material of the human eye \cite{Mb3}. The more complex simulations include vacuum (G4\_Galactic) and various shielding materials (i.e., aluminium, kevlar). One of the models also includes human tissue (G4\_BONE\_CORTICAL\_ICRP, G4\_BRAIN\_ICRP). 

The physics list manages the physics processes in the entire simulation. We use a physics list named FTFP\_BERT for our simulations. This physics list contains a hadronic component, an electromagnetic component, a decay component, and a neutron tracking cut \cite{Mo5}. The hadronic component includes elastic, inelastic, and capture processes. The hadronic component uses many different models to deal with various hadronic processes, such as the Fritiof parton model (FTF), the Bertini, and Precompound models to simulate inelastic processes, the G4ChipsElasticModel and the G4HadronElastic model for elastic processes, and the G4NeutronRadCapture model, the G4MuonMinusCap- ture, the BertiniCaptureAtRest, and the FritiofCaptureAtRest, to maintain the capture and stopping processes in the simulation \cite{Mo5}. The electromagnetic component, the EM Opt4, was manually added to the physics list. The EM Opt4, or \\G4EmStandardPhysics\_option4, consists of physics dealing with electromagnetic processes, such as the BetheHeitler5D model for electron/positron pair production, the Klein--Nishina model for Compton scattering, or the Livermore models for the photo-electric effect and Rayleigh scattering \cite{Mo6}. The G4Decay processes handle the decay component of the simulations \cite{Mo5}. The FTFP\_BERT physics list is the recommended choice for cosmic ray applications.

The OpNovice example from Geant4's extended optical examples demonstrates how to incorporate optical processes into a Geant4 simulation. It includes Cherenkov radiation, scintillation, Rayleigh scattering, Mie scattering, absorption, and boundary processes. Our model uses all these processes except scintillation because water is considered an extremely weak scintillator. Geant4 provides these processes as predefined classes within G4VProcess for user convenience. G4Cerenkov handles Cherenkov radiation, G4OpRayleigh handles Rayleigh scattering, G4OpMieHG handles Mie scattering, G4OpAbsorption handles absorption, and G4OpBoundaryProcess handles boundary processes.

The simplest Geant4 models consists of two concentric objects. Four of the models belong to this category. The first model (for simplicity, called \textbf{CG-MA} (\textbf{C}ube geometry \textbf{G}eant4 - \textbf{M}ore \textbf{A}ir)) consists of two concentric blocks. The outer block, filled with air (a user-programmed material Air comprising 70\% nitrogen and 30\% oxygen), was 10 cm in the $x$-direction, 15 cm in the $y$-direction, and 10 cm in the $z$-direction. In the centre of the Air block is a smaller (inner) Water block, filled with water (user-programmed material consisting of hydrogen and oxygen in a ratio of 2:1). It extends 2 cm in the  $x$-direction, 3 cm in the $y$-direction, and 2 cm in the $z$-direction . The smaller water block uses these specific dimensions due to their similarity in size to the human eye. According to medical studies, the dimensions of the human eye are approximately 23.7 mm along the $x$-direction (sagittal), 22.0--24.8 mm along the $y$-direction (axial), and 24.2 mm  along the $z$-direction (transversal) \cite{Mb2}.  

The second model (for easier reference, called \textbf{CG-LA} (\textbf{C}ube geometry \textbf{G}eant4 - \textbf{L}ow \textbf{A}ir) was the answer to our question whether the amount of air around the water block influences the results. Therefore, we reduced the value of the extents along the $x$-, $y$-, and $z$-directions of the Air block. The new dimensions were only 2 mm bigger than the dimensions of the Water block (thus, $x = 2.2$ cm, $y = 3.2$ cm, and $z = 2.2$ cm).

The third and fourth model also mirror the geometry of the previous models; however, the inner geometry changed from a block to a sphere filled with water. Its diameter changes depending on the model. The diameter in the third model is 2.0 cm (called \textbf{SG-2.0} (\textbf{S}phere geometry \textbf{G}eant4 \textbf{- 2.0} diameter)) to complement and cross-check the previous models. However, to accurately simulate the size of the actual human eye, the fourth model uses a sphere with diameter of 2.5 cm (model name \textbf{SG-2.5} (\textbf{S}phere geometry \textbf{G}eant4 \textbf{- 2.5} diameter)).

In the more complex models the geometry slightly changes. In this category we have further six models. The first, (called \textbf{SG-Vac} (\textbf{S}phere geometry \textbf{G}eant4 \textbf{- Vac}uum)) added a block with vacuum (G4\_Galactic) around the air block. The dimensions of the vacuum block are: $x = 2.4$ m, $y = 2.4$ m, $z = 2.4$ m. The air block was significantly increased to $x = 2$ m, $y = 2$ m, $z = 2$ m. The innermost object is a sphere filled with water with diameter of 2.5 cm. 

Next two models include shielding materials. The first one contains aluminium (called \textbf{SG-Al} (\textbf{S}phere geometry \textbf{G}eant4 \textbf{- Al}uminium)). This model has four concentric objects. First is the vacuum block (2.4 m x 2.4 m x 2.4 m), inside of which is the shielding layer, a block filled with aluminium (2.01 m x 2.01 m x 2.01 m). It wraps around the air block (with dimensions: 2 m x 2 m x 2 m). Inside the air block is the water sphere simulating the human eye (diameter: 2.5 cm). The second model contains both aluminium and kevlar (G4\_KEVLAR) as a shielding materials (model name \textbf{SG-ISS} (\textbf{S}phere geometry \textbf{G}eant4 \textbf{- ISS})). It simulates the situation on the orbit (specifically on ISS). The geometry consists of five concentric layers. The first, and outermost, layer is the vacuum layer (with dimensions: 2.4 m x 2.4 m x 2.4 m). The second layer is the aluminium block (dimensions: 2.02 m x 2.02 m x 2.02 m). Right after is the kevlar layer (its dimensions are: 2.01 m x 2.01 m x 2.01 m). Next is the air block (2 m x 2 m x 2 m) inside of which resides the water sphere with diameter of 2.5 cm. 

The last three models simulate different areas of human body. The first two models simulate human head, while the last model simulates the retina of the eye. The first two models that simulate human head differ in the materials used in the simulation. The first one (named \textbf{SG-H\_W} (\textbf{S}phere geometry \textbf{G}eant4 \textbf{-H}ead\textbf{\_W}ater)) uses water to approximate the human tissue. It consists of three concentric objects: the outermost vacuum layer (2.4 m x 2.4 m x 2.4 m), the middle layer filled with air (2 m x 2 m x 2 m), and the innermost spherical layer filled with water, whose diameter was set to 22 cm to better approximate the size of the human head. The second model (model name \textbf{SG-H\_HT} (\textbf{S}phere geometry \textbf{G}eant4 \textbf{-H}ead \textbf{\_H}uman \textbf{T}issue)) consists of four concentric objects: the outermost is the vacuume layer (2.4 m x 2.4 m x 2.4 m), the second layer is the air block (2 m x 2 m x 2 m), and the last two layers are the bone layer (G4\_BONE\_CORTICAL\_ICRP) with diameter of 22 cm, and the brain layer (G4\_BRAIN\_ICRP) with diameter of 20.8 cm. 

The last model simulates the retina on the backside of the eye. The model (named \textbf{CG-Ret} (\textbf{C}ube geometry \textbf{G}eant4 \textbf{-Ret}ina)) consists of two concentric layers: the vacuum layer (dimensions: 2.4 m x 2.4 m x 2.4 m), and the inner block filled with water with dimensions: $x = 1.5$ cm, $y = 1.5$ cm, and $z = 0.3$ mm, simulating the retina at the back of the eye. 

The simulation initially uses protons of various energies as the primary particles, i.e. the particles injected into the material. We use protons as the primary particles because a large portion of CRs consists of hydrogen nuclei (thus protons). Later, heavier elements, from helium to oxygen and additional iron, were simulated. The shape of the beam of primary particles is a point source, and the particles travel along the $z$-axis. The proton simulations use energies from 1 MeV to 50 GeV, mainly 1, 10, 100, 500 MeV, and 1, 2, 3, 5, 10, 30 and 50 GeV. We chose this energy interval to simulate the range of energies in the CR spectrum with the highest intensities at 1 AU.

To better understand the interaction of the CR particles with the human eye, we also simulate muons. Muons were selected due to the persistent question of why we do not see the LFs down at the Earth's surface. As one of the particles created during the CR particles' interaction with the Earth's atmosphere, muons are the best candidates for the cause of LFs at the Earth's surface. Both of these primary particles were given energies from 1 MeV to 10 GeV.

Geant4 provides information on all particles in the simulation, including primary particles, for example, their PDG code, kinetic energy, momentum, and other properties. Simulations use the SteppingAction class to record all pertinent data from step 1 until the final step into the root and CSV files. There is no way to record step 0 through the SteppingAction; therefore, we implemented the TrackingAction class to keep track of step 0 for all particle types. TrackingAction takes data from step 0 and records them in the root and CSV files. Step 0 is valuable information for our analysis because it carries the information about the process that created the secondary particle, for example, Cherenkov radiation.

\begin{figure}[ht!]
\centering
\noindent\includegraphics[scale = 0.3]{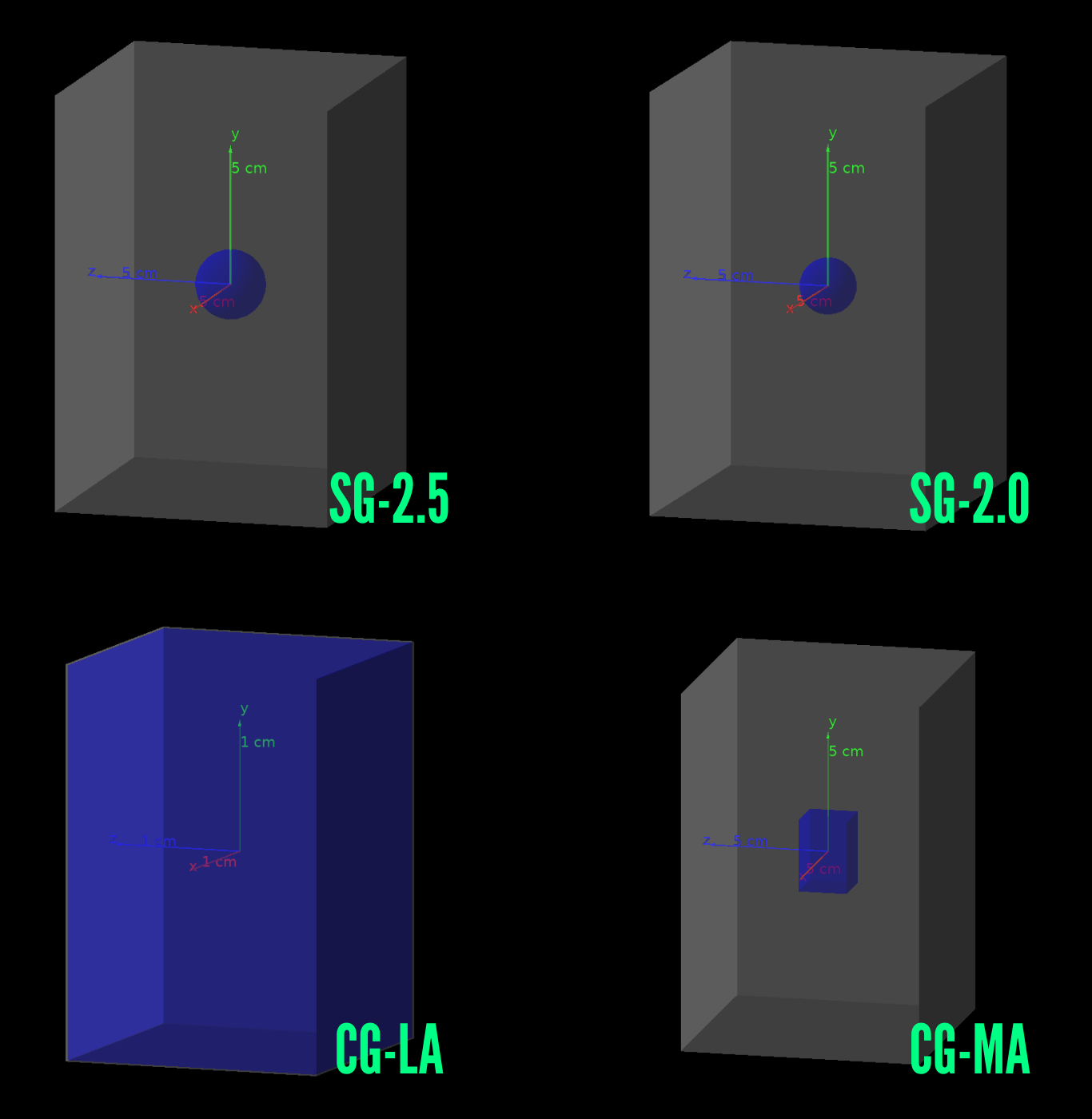}
\caption{Geometries of Human-eye water models}
\label{fig:Geo_model}
\end{figure}

Table \ref{tab.2} provides an overview of the various models of the human eye. It contains details about the different models in Geant4, such as their geometry, material and dimensions, type, and number of primary particles. Figure \ref{fig:Geo_model} shows the geometry of different models of the human eye. Four of the Geant4 models are depicted on Figure \ref{fig:Geo_model}. 

\begin{table}
 \caption{Versions of the model of the human eye (G-Geant4, C-Cube, S-Sphere)}
 \label{tab.2}
 \centering
    \begin{tabular}{|c|ccc|c|}
        \hline
        Model version & & Geometry & & Primary particle \\
        & shape & material & dimensions [cm] & /number of p \\
        \hline
        & Block & air & 10 x 15 x 10 & proton/10 000$^{\ast}$ \\
        CG-MA & Block & water & 2 x 3 x 2 & alpha/10 000$^{\ast}$ \\
        & & & & muon/10 000\\
        \hline
        CG-LA & Block & air & 2.2 x 3.2 x 2.2 & proton/10 000$^{\ast}$ \\
        & Block & water & 2 x 3 x 2 & \\
        \hline
        SG-2.0 & Block & air & 10 x 15 x 10 & proton/10 000$^{\ast}$ \\
        & Sphere & water & d = 2 & \\
        \hline
        SG-2.5 & Block & air & 10 x 15 x 10 & proton/10 000$^{\ast}$ \\
        & Sphere & water & d = 2.5 & \\
        \hline
        & Block & G4\_Galactic & 240 x 240 x 240 & proton/10 000$^{\ast}$ \\
        SG-Vac & Block & air & 200 x 200 x 200 & antimuon/10 000$^{\ast}$\\
        & Sphere & water & d = 2.5 & He--O/1000$^{\ast}$ \\
        & & & & iron (Fe)/1000$^{\ast}$ \\
        \hline
        & Block & G4\_Galactic & 240 x 240 x 240 & proton/10 000$^{\ast}$ \\
        SG-Al & Block & Al & 201 x 201 x 201 & proton/10 000$^{\dagger}$ \\
        & Block & air & 200 x 200 x 200 & \\
        & Sphere & water & d = 2.5 & \\
        \hline
        & Block & G4\_Galactic & 240 x 240 x 240 & proton/10 000$^{\ast}$\\
        & Block & Al & 202 x 202 x 202 & proton/10 000$^{\dagger}$ \\
        SG-ISS & Block & G4\_Kevlar & 201 x 201 x 201 & lithium/10 000$^{\dagger}$\\
        & Block & air & 200 x 200 x 200 & iron/1000$^{\ast}$ \\
        & Sphere & water & d = 2.5 & \\
        \hline
        & Block & G4\_Galactic & 240 x 240 x 240 & \\
        SG-H\_W & Block & air & 200 x 200 x 200 & lithium/1000$^{\ast}$ \\
        & Sphere & water & d = 22.0 & \\        
        \hline
        & Block & G4\_Galactic & 240 x 240 x 240 & \\
        SG-H\_HT & Block & air & 200 x 200 x 200 & lithium/1000$^{\ast}$ \\
        & Sphere & G4\_Bone\_Cortical\_ICRP & d = 22.0 & \\
        & Sphere & G4\_Brain\_ICRP & d = 20.8 & \\
        \hline
        CG-Ret & Block & G4\_Galactic & 240 x 240 x 240 & proton/1000$^{\ast}$\\
        & Block & water & 1.5 x 1.5 x 0.03 & lithium/1000$^{\ast}$ \\
        \hline
        \multicolumn{5}{l}{$^{\ast}$ point source; $^{\dagger}$ circle source} \\
    \end{tabular}
\end{table}

\section{Results}

In the previous analysis, where Cherenkov light production was not evaluated \cite{Rb1, Rb2}, we showed that the nuclear interactions of primary protons with the water eye model did not produce visible light. The resulting gamma-ray spectra also has a low energy part, which could reach the energies of visible light, but in negligible amounts. 

\begin{figure}[ht!]
\noindent\includegraphics[width=\textwidth]{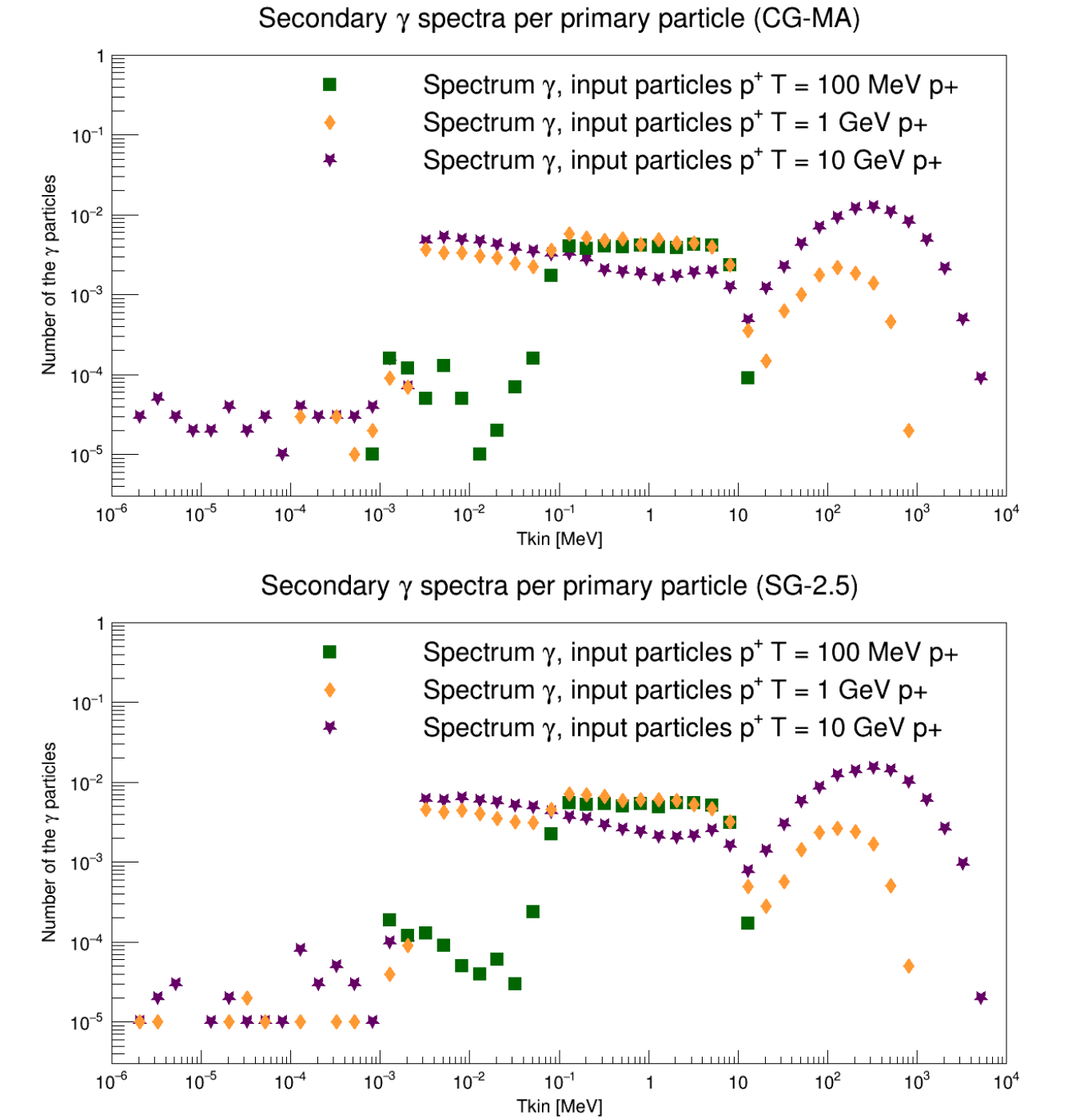}
\caption{Spectrum of gamma rays produced by 100 MeV, 1 GeV and 10 GeV primary protons in Geant4 CG-MA simulation setup (topmost graph) and in SG-2.5 setup (lower graph) without the Cherenkov light production}.
\label{fig:Ekin_vs_gamma_intensities}
\end{figure}

The Geant4 simulations, performed with 100 000 primary particles in the CG-MA setup and presented in Figure \ref{fig:Ekin_vs_gamma_intensities}, show the spectra for 100 MeV, 1 GeV, and 10 GeV protons. No visible photons (in the eV range) were produced at 100 MeV or 1 GeV, while 10 GeV protons generated only a few, requiring more than one thousand primary protons to yield a single photon.

In the following, we present in detail the simulations with added Cherenkov light production.

\subsection{Results of the Geant4 simulations}

\begin{figure}[ht!]
\noindent\includegraphics[width=\textwidth]{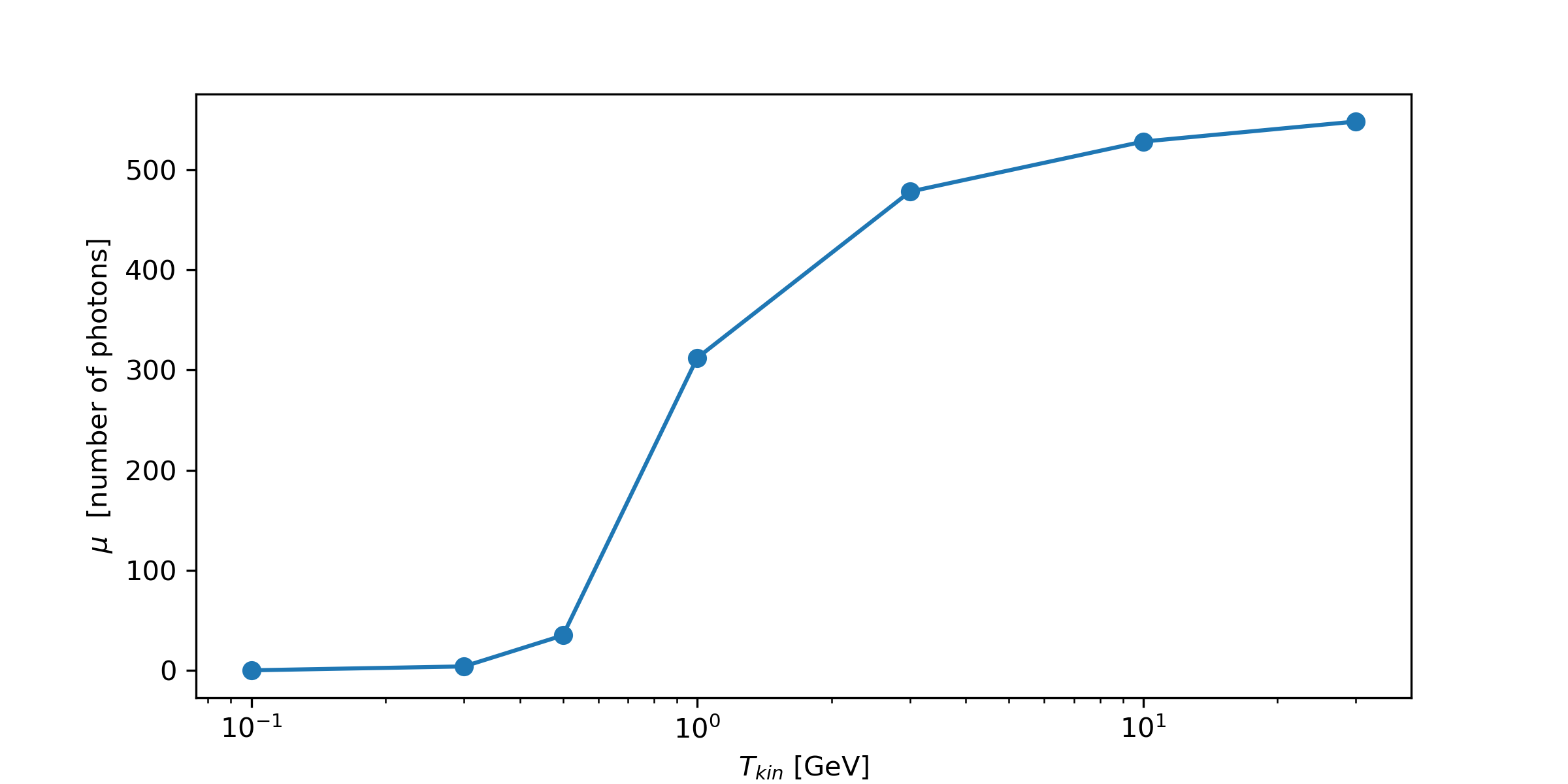}
\caption{Mean value $\mu$ of number of produced photons (wavelength range 380--700 nm) as a function of the kinetic energy of the primary protons. Results from Geant4 CG-MA simulation setup.}
\label{fig:Ekin_vs_mu_all_energies}
\end{figure}

In Figure \ref{fig:Ekin_vs_mu_all_energies} we present the mean values $\mu$ of the number of created photons (wavelength range 380--700 nm) per primary proton in our simulation setting CG-MA as a function of the primary proton kinetic energy. The Geant4 simulation provides mean values $\mu$ increasing from values close to zero around the expected energy 450--460 MeV, through 312 photons for 1 GeV primary protons, to a plateau of nearly 550 photons over 10 GeV. The refractive index of water has, in Geant4 Water G4MaterialPropertiesTable, values in the range 1.3435--1.3608, for a range of photon energies of 1.771--3.263 eV (wavelength range 700--380 nm). For a water refractive index equal to 1.35, the speed of light in water equals 0.7407$c$. A proton with kinetic energy 458 MeV has a speed very close to the  0.7401$c$. This kinetic energy sets the approximate threshold over which a proton in water has a velocity higher than the speed of light and creates Cherenkov light. 

The current simulation setup includes the production of Cherenkov photons. When we switched off this production, we got results without optical photons. Optical photons are created only by Cherenkov emission.

Depending on the position at Earth orbit, the CR proton intensities vary. A proton with kinetic energy higher than 450 MeV will collide with/pass 1 cm$^2$ from solid angle 1 steradian a few times per minute in regions with the lowest cutoff rigidity. The total flux of protons at the boundary of the magnetosphere varies during the solar cycle in the energy range from 432 MeV to nearly 100 GeV between approximately 10 to 20 protons per cm$^2$ minute and steradian \cite{Rb3}. A proton with energy higher than 100 GV hits the/passes the 1 cm$^2$ from solid angle one steradian once per 80 minutes (flux approximately 2 protons per m$^2$ s sr). Because of the low proton flux over 100 GeV, and the threshold for the creation of Cherenkov light ($\approx$ 458 MeV), we approximate the flux of CRs reaching the astronauts in space close to Earth by a proton flux in a rigidity range from 1 to 100 GV (kinetic energy from 433 MeV to 99.07 GeV). As a result, in the most exposed parts of Earth's orbit, an astronaut's eye faces approximately a hundred protons per minute (from 2$\pi$ sphere, i.e. from the sky above) with energy over the speed of light in water. 

The presented mean values $\mu$ of the number of created photons (wavelength range 380--700 nm) are evaluated from 10 000 interacting primary protons. Every one of them produces a similar, but not the same number of photons in the eye water block model. Unless stated otherwise, the term photon in the following text refers only to photons in the visible wavelength range (380–700 nm), corresponding to photons detectable by the human eye.

\begin{figure}[!htbp]
\noindent\includegraphics[width=\textwidth]{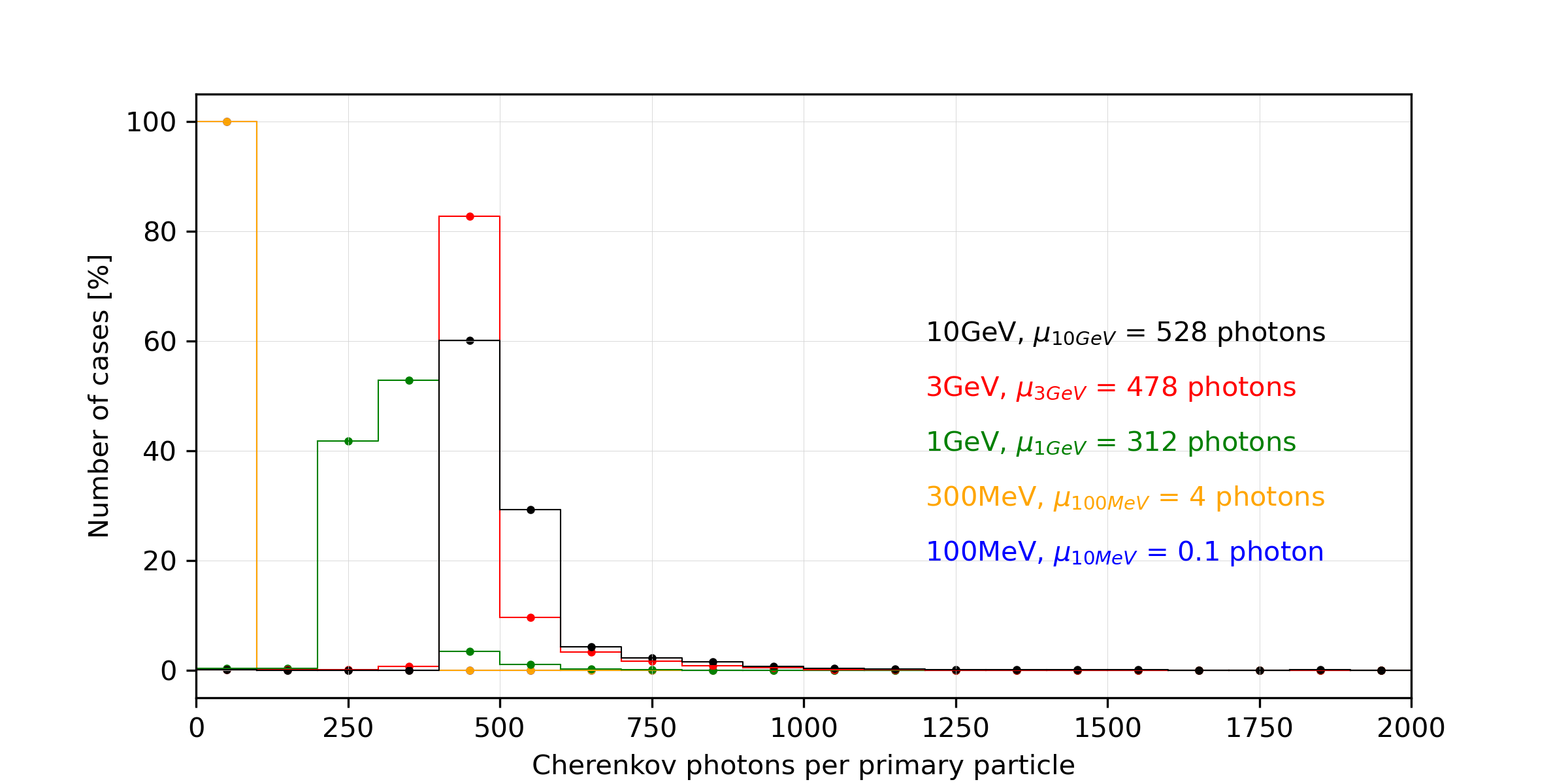}
\caption{Histograms of number of photons (wavelength range 380--700 nm) created in test eye water block Geant4 model CG-MA by primary protons with energies 100 MeV, 300 MeV, 1 GeV, 3 GeV and 10 GeV.}
\label{fig:histogram_of_created_cherenkov_photons_for_selected_energies}
\end{figure}

In Figure \ref{fig:histogram_of_created_cherenkov_photons_for_selected_energies} we show the distributions of the number of photons created in the water block CG-MA. For the kinetic energy of primary proton 1 GeV (green histogram) with $\mu$ 312 photons, we see that in more than 95\% of the cases the number of created Chrenkov photons is in the range 200--400 photons. The histogram becomes wider with increasing kinetic energy of the primary protons, for 3 GeV this becomes 83\% of the cases between 400 and 500 photons, with 10\% of the cases creating between 500 and 600 photons. For a 10 GeV primary proton, 60\% of the cases are in the range 400--500 photons and 29\% between 500 and 600 photons. For a 30 GeV primary proton (case not shown in Figure \ref{fig:histogram_of_created_cherenkov_photons_for_selected_energies}) the situation is similar to that of the 10 GeV histogram, with 56\% of the cases between 400 and 500 photons, and 33\% of cases between 500 and 600 photons. The histograms show that cosmic ray protons with kinetic energies of a few GeV, where the cosmic ray spectrum is close to its maximum, produce several hundred Cherenkov photons in the water block model of the human eye.

\begin{figure}[!htbp]
\centering
\includegraphics[width=0.98\textwidth]{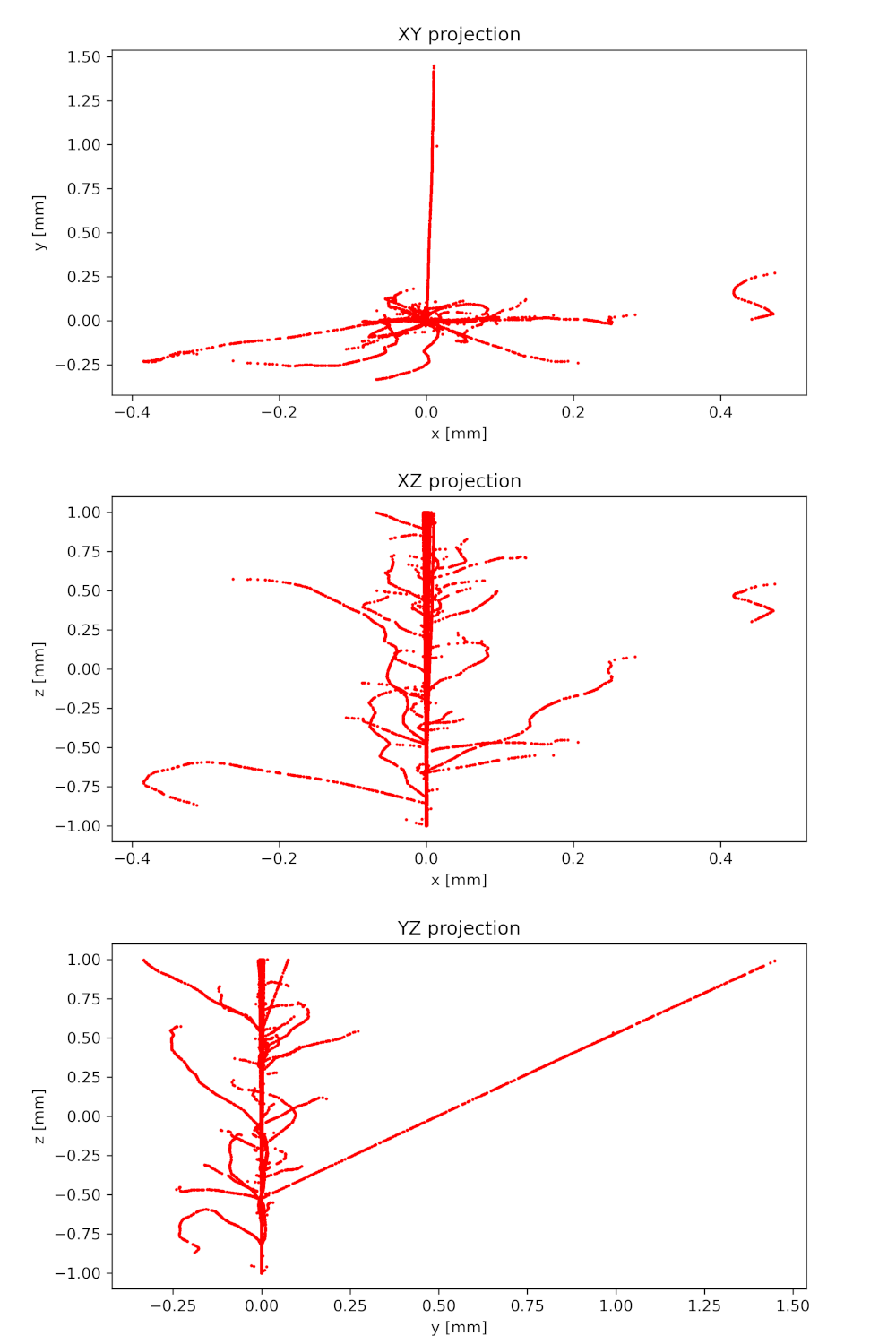}
\caption{The points of creation of 500 thousand photons (wavelength range 380--700 nm), from 1 GeV primary protons.}
\label{fig:Created_photons_1GeV_primary_example_5e6_photons}
\end{figure}

The primary protons are injected along the $z$-axis of the CG-MA setup. The protons pass through the whole water block in most cases, very close to the $z$-axis. Along this primary proton trajectory, Cherenkov photons are created. In Figure \ref{fig:Created_photons_1GeV_primary_example_5e6_photons} we show an example of the points where photons were created. The points of creation of the first 500 thousand photons, from approximately the first one thousand simulated primary protons with injection kinetic energy 1 GeV, are shown as red dots. We can see points of creation following the primary proton trajectories. The red points show that most primary protons stay close to the $z$-axis and a few of them are scattered. Then most of the photons are created close to the $z$-axis. In the figure, we show the $xz$, $yz$, and $xy$ projections of the simulations.

\begin{figure}[!htbp]
\noindent\includegraphics[width=\textwidth]{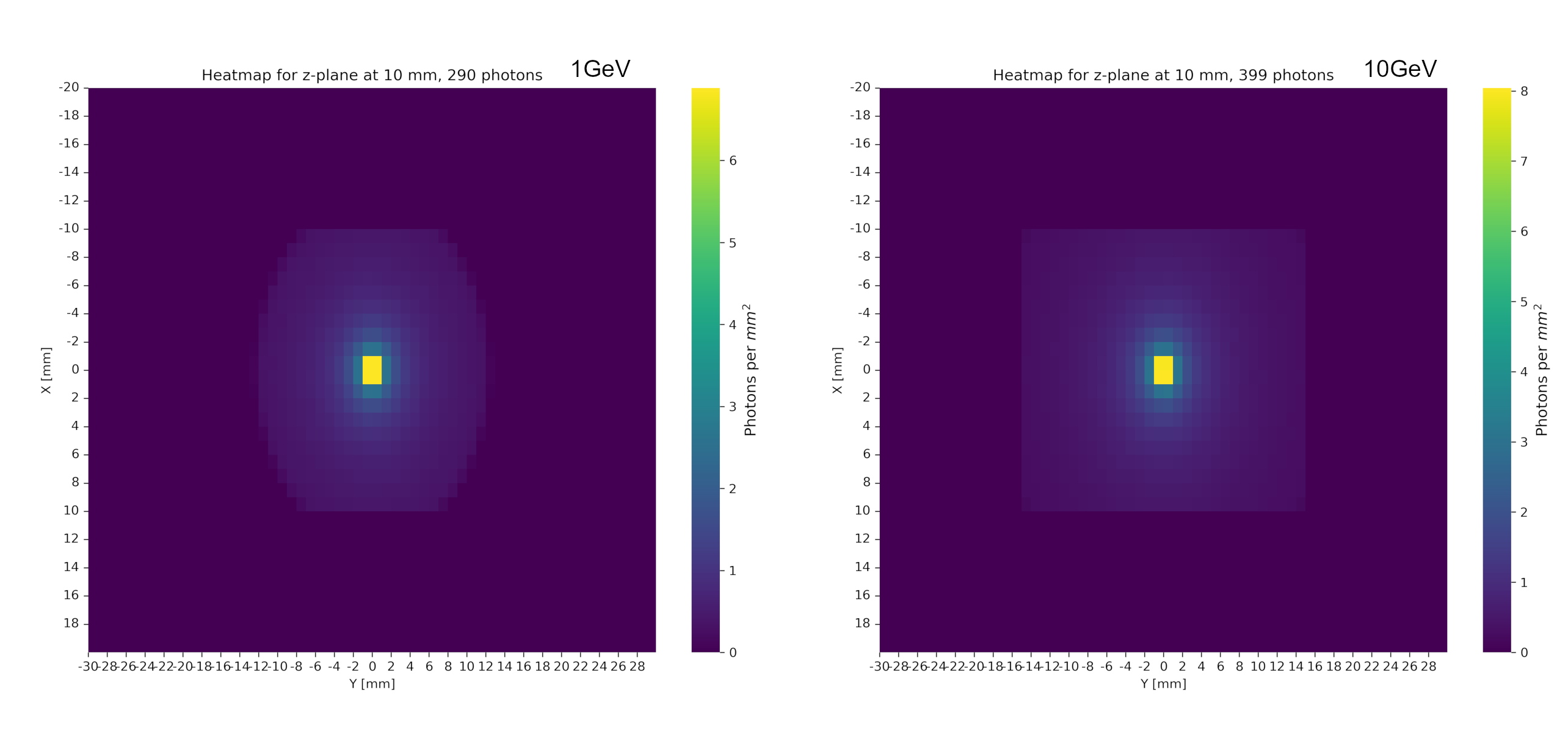}
\caption{Heat map of distribution of photons in the last layer of the eye water box model CG-MA simulated in Geant4  for 1 and 10 GeV primary proton.}
\label{fig:XY_heatmap_10GeV_z_plane_20_crosses_1GeV_and_10GeV_primary_proton}
\end{figure}

We trace trajectories of all created Cherenkov photons and evaluate if and where their trajectory crosses different $xy$ layers in the water block. The last layer of water block was crossed by 290 photons for 1 GeV primary protons and 399 photons for 10 GeV primary protons. The heat map of their distribution is presented in Figure \ref{fig:XY_heatmap_10GeV_z_plane_20_crosses_1GeV_and_10GeV_primary_proton}. The flux of photons close to the central $z$-axis, through mm$^2$ pixels, is close to 6 photons per mm$^2$ for 1 GeV proton and approximately 8 photons per mm$^2$ for 10 GeV proton and decreases quickly with distance from the $z$-axis. In a central area of 16 mm$^2$, defined by a square $\pm$ 2 mm from the $z$-axis, we have 56 photons for a 1 GeV primary proton. In a central area of 36 mm$^2$ $\pm$ 3 mm from $z$-axis, it is 84 photons, and in an area of 64 mm$^2$ $\pm$ 4 mm from the $z$-axis, 112 photons. In a central area of 1 cm$^2$, it is 140 photons. 
For a 10 GeV proton, the central 16 mm$^2$ has 66 photons and the central 1 cm$^2$ 167 photons. Table \ref{fig:along-Y_axis-last_Z-layer-table} shows 
the number of photons per mm$^2$ along the $y$ axis, from positions within the range -15 mm to 15mm from the centre of the box, and for $z$ = 10 mm (i.e. through last layer of the water block CG-MA) for 1 GeV and 10 GeV primary particles. The presented number of photons per mm$^2$ in Table \ref{fig:along-Y_axis-last_Z-layer-table} shows that the angle of the Cherenkov light column centred around the direction of the primary particles changes the illuminated area with the energy of primary particles in a minimal way. The number of photons crossing the last layer  of the water box is similar for 1 GeV and 10 GeV primary protons.

\begin{table}[ht!]
\centering
\caption{Number of photons (wavelength range 380--700 nm) per mm$^2$ along $y$-axis at $z$ = 10 mm (i.e. crossing last layer) for 1 GeV and 10 GeV particles.}
\label{fig:along-Y_axis-last_Z-layer-table}
\begin{tabular}{|c|c|c|}
\hline
\textbf{Y position } & \textbf{1 GeV} & \textbf{10 GeV}  \\ \hline
\textbf{[mm]} & \textbf{photons per mm$^2$} & \textbf{photons per mm$^2$}  \\ \hline
-14.5 & 0.0044 & 0.316 \\ \hline
-10.5 & 0.377 & 0.441 \\ \hline
-8.5 & 0.476 & 0.540 \\ \hline
-6.5 & 0.625 & 0.722 \\ \hline
-4.5 & 0.874 & 1.061 \\ \hline
-2.5 & 1.588 & 1.888 \\ \hline
-0.5 & 6.861 & 8.023 \\ \hline
0.5 & 6.862 & 8.042 \\ \hline
1.5 & 2.569 & 3.087 \\ \hline
3.5 & 1.150 & 1.371 \\ \hline
5.5 & 0.728 & 0.864 \\ \hline
7.5 & 0.523 & 0.629 \\ \hline
9.5 & 0.426 & 0.466 \\ \hline
14.5 & 0.0049 & 0.319 \\ \hline
\end{tabular}
\end{table}

The results show that most of the Cherenkov protons are created inside the eye following the direction of the incoming primary particle. The result is a narrow cone of visible photons affecting a relatively small area on the back side of the eye. This result is consistent with the observation of LFs by astronauts, where most of them have a spot or starlike (66\%, see Introduction for details) appearance.

To check the setup of the Geant4 CG-MA simulation, we made the air region around the water block minimal, with the boundary of the whole setup CG-LA just a millimeter around the water block. Details about the configurations were mentioned in the previous chapter. The results for 1 GeV primary protons are almost identical to the default setup with a wider region of air around the water block. The simulation for 10 000 primary protons yields a mean value of created photons 310. This is almost the same as the 312 photons for the default setup CG-MA. In conclusion, the size of the box of air around the water box only negligibly affects the simulation results. This confirms that we could use the default simulation setup CG-MA results in our analysis.

\subsubsection{Spherical eye model}

As our investigation of light flash mechanisms progressed, the eye model was refined to better represent the anatomical geometry. The results presented in the previous sections were obtained using a simplified CG-MA cubic eye model, which facilitated the initial evaluation of particle interactions and light production. In the following sections, however, we employ a more realistic spherical eye model with a diameter of 2.0 cm and 2.5 cm, setup SG-2.0 and SG-2.5, to assess the influence of eye geometry on the predicted light-flash characteristics.

The mean value of the number of created photons for a 1 GeV primary proton was 311 for the setup with a radius of 1.0 cm, referred to as SG-2.0.  This is the same number as for the water box model CG-MA. The trajectory of the primary particles mostly followed the direction along the $z$-axis and  had the same length, 2 cm, in both eye models.  This shows that due to the sameness of the lengths of the trajectories of the primary particles and the emission of the Cherenkov photons mostly along the $z$-axis, the sphere and box eye models provide very similar results. The number of created photons in SG-2.0 were, in 95\% of the cases, in the range between 200 and 400 photons, again very similar to the CG-MA model.  

For a 10 GeV primary proton, the mean number of created photons is 522 in the SG-2.0 eye model and 662 photons in the SG-2.5 eye model. Again, for SG-2.0, this is almost the same as the 528 photons of the water box CUBE\_ GMA eye model (the numbers of photons are rounded to whole numbers). The most cases, 90\%, there are  between 400 and 600 photons for SG-2.0, similar to the CG-MA eye model. 

One further refinement of the eye model is to adjust its characteristic size to more closely represent the average anatomical dimensions of the human eye. In the case of a model of the human eye in \cite{Mb2} it is a radius of 1.25 cm instead of 1.0 cm. This increases the length of the trajectory of a primary particle in a sphere model from 2 cm to 2.5 cm. We tested this new geometry, referred to as SG-2.5, and the mean value of the number of created photons for a 1 GeV proton is 388, 25\% higher than in the smaller spherical model. The number of photons created by a 10 GeV primary particle is 661, 27\% higher than in the case with a 2-cm water sphere.

Cherenkov emission depends on the length of the trajectory of the particle in the material, which is comparable for both simulation configurations: the number of created photons is very similar for the water box and the water sphere. The comparison between the cubic and spherical geometries shows that the dominant contribution to the photon yield is determined by the charged particle track length in the medium, while the detailed shape of the outer boundary has only a secondary influence

The conclusion about the sphere water box setup is that this more realistic eye model does not significantly change the results from the previous model, however, in the next simulations presented in the article, we use a sphere SG-2.5 based setup to keep the simulation closer to human physiology.

\subsection{Cherenkov light production model}

A Cherenkov light model was developed and used for simulations that could not be fully performed in the Geant4 package due to performance and disk space limitations. The model uses the Frank–Tamm formula and assumes a water refractive index of 1.35. The particle track length inside the eye water model, approximated as a water sphere with a radius of 1.25 cm, was set to 4r/3 (1.667 cm), which represents the average chord length in a sphere for isotropically incoming particles. Cherenkov light production was evaluated in visible range from 380 till 700 nm for the first 30 elements of the periodic table (from hydrogen to zinc), using a 10 MeV energy step from 0.01 GeV up to 500 GeV.

\begin{figure}[!htbp]
\noindent\includegraphics[width=\textwidth]{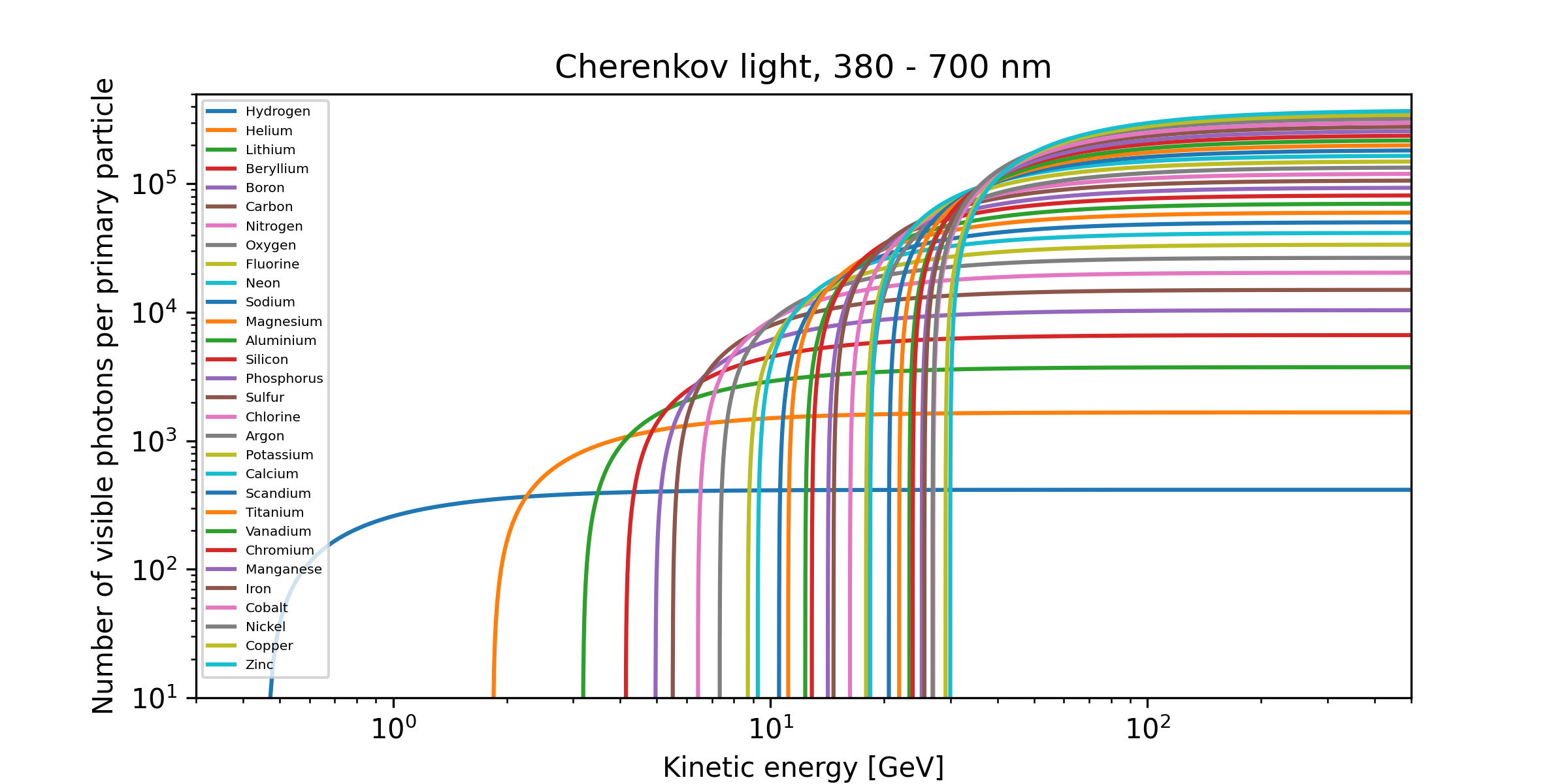}
\caption{Cherenkov light production dependence on the kinetic energy of the whole nucleus for the first 30 elements of the periodic table. Evaluated for a 1.667 cm long trajectory in water, representing the average length of nuclei trajectories in a 2.5 cm water eye model.}
\label{fig:Cherenkov_light_simple_model}
\end{figure}

Figure \ref{fig:Cherenkov_light_simple_model} shows the dependence of Cherenkov light production on the particle type for elements from hydrogen to zinc, in the visible wavelength range, up to 500 GeV. The figure illustrates that the number of photons produced in the eye water model strongly depends on the nucleus, from a few hundred photons for a hydrogen nucleus (proton) to more than three hundred thousand for a zinc nucleus, approximately a thousand times more than for a proton.
After the threshold energy $T_{kin,Ch}$, where the particle exceeds the speed of light limit in water, the photon yield increases steeply and eventually reaches a plateau. Table \ref{tab simple model} summarizes the threshold energy at which Cherenkov production begins $T_{kin,Ch}$, the energy at which the plateau appears, and the approximate plateau intensity for all 30 evaluated elements at start of plateau and at the 500 GeV. The onset of the plateau was defined as the energy at which the slope of the Cherenkov photon yield for a given element becomes smaller than 0.01 \% per GeV.

\begin{table}
 \caption{Table showing the dependence of Cherenkov light production on the particle type for elements from Hydrogen to Zinc (see details in the text).}
 \label{tab simple model}
 \centering
    \begin{tabular}{|c|c|c|c|c|c|c|}
        \hline
\textbf{Element} & \textbf{Atomic} & \textbf{$T_{kin, Ch}$} & \textbf{Plateau start} & \textbf{N. of photons } & \textbf{N. of photons} \\ 
 & \textbf{number} & \textbf{$[GeV]$} & \textbf{[GeV]} & \textbf{plateau start} & \textbf{at 500GeV} \\ 
        \hline
Hydrogen & 1 & 0.462 & 27.01 & 415.2 & 415.7 \\
        \hline
Helium & 2 & 1.835 & 66.47 & 1657.2 & 1662.8 \\
        \hline
Lithium & 3 & 3.181 & 94.97 & 3722.8 & 3740.9 \\
        \hline
Beryllium & 4 & 4.131 & 112.44 & 6612.0 & 6649.6 \\
        \hline
Boron & 5 & 4.955 & 126.42 & 10323.3 & 10388.4 \\
        \hline
Carbon & 6 & 5.506 & 135.3 & 14858.3 & 14957.7 \\
        \hline
Nitrogen & 7 & 6.421 & 149.35 & 20208.0 & 20354.9 \\
        \hline
Oxygen & 8 & 7.334 & 162.66 & 26374.6 & 26579.7 \\
        \hline
Fluorine & 9 & 8.709 & 181.59 & 33344.9 & 33626.2 \\
        \hline
Neon & 10 & 9.25 & 188.74 & 41150.1 & 41506.4 \\
        \hline
Sodium & 11 & 10.539 & 205.14 & 49745.5 & 50199.4 \\
        \hline
Magnesium & 12 & 11.141 & 212.56 & 59176.4 & 59727.4 \\
        \hline
Aluminium & 13 & 12.368 & 227.21 & 69392.4 & 70060.8 \\
        \hline
Silicon & 14 & 12.874 & 233.09 & 80451.8 & 81235.6 \\
        \hline
Phosphorus & 15 & 14.198 & 248.08 & 92276.4 & 93196.7 \\
        \hline
Sulfur & 16 & 14.696 & 253.58 & 104957 & 106011 \\
        \hline
Chlorine & 17 & 16.25 & 270.31 & 118373 & 119578 \\
        \hline
Argon & 18 & 18.312 & 291.6 & 132547 & 133899 \\
        \hline
Potassium  & 19 & 17.923 & 287.65 & 147717 & 149225 \\
        \hline
Calcium & 20 & 18.372 & 292.2 & 163632 & 165301 \\
        \hline
Scandium & 21 & 20.608 & 314.26 & 180174 & 181983 \\
        \hline
Titanium & 22 & 21.942 & 327.0 & 197597 & 199544 \\
        \hline
Vanadium & 23 & 23.352 & 340.13 & 215803 & 217874 \\
        \hline
Chromium & 24 & 23.835 & 344.56 & 234916 & 237145 \\
        \hline
Manganese & 25 & 25.184 & 356.75 & 254719 & 257052 \\
        \hline
Iron & 26 & 25.599 & 360.46 & 275445 & 277936 \\
        \hline
Cobalt & 27 & 27.015 & 372.91 & 296824 & 299382 \\
        \hline
Nickel & 28 & 26.905 & 371.95 & 319236 & 321999 \\
        \hline
Copper & 29 & 29.13 & 391.06 & 342063 & 344754 \\
        \hline
Zinc & 30 & 29.97 & 398.13 & 365908 & 368665 \\
        \hline
    \end{tabular}
\end{table}

\begin{figure}[!htbp]
\noindent\includegraphics[width=\textwidth]{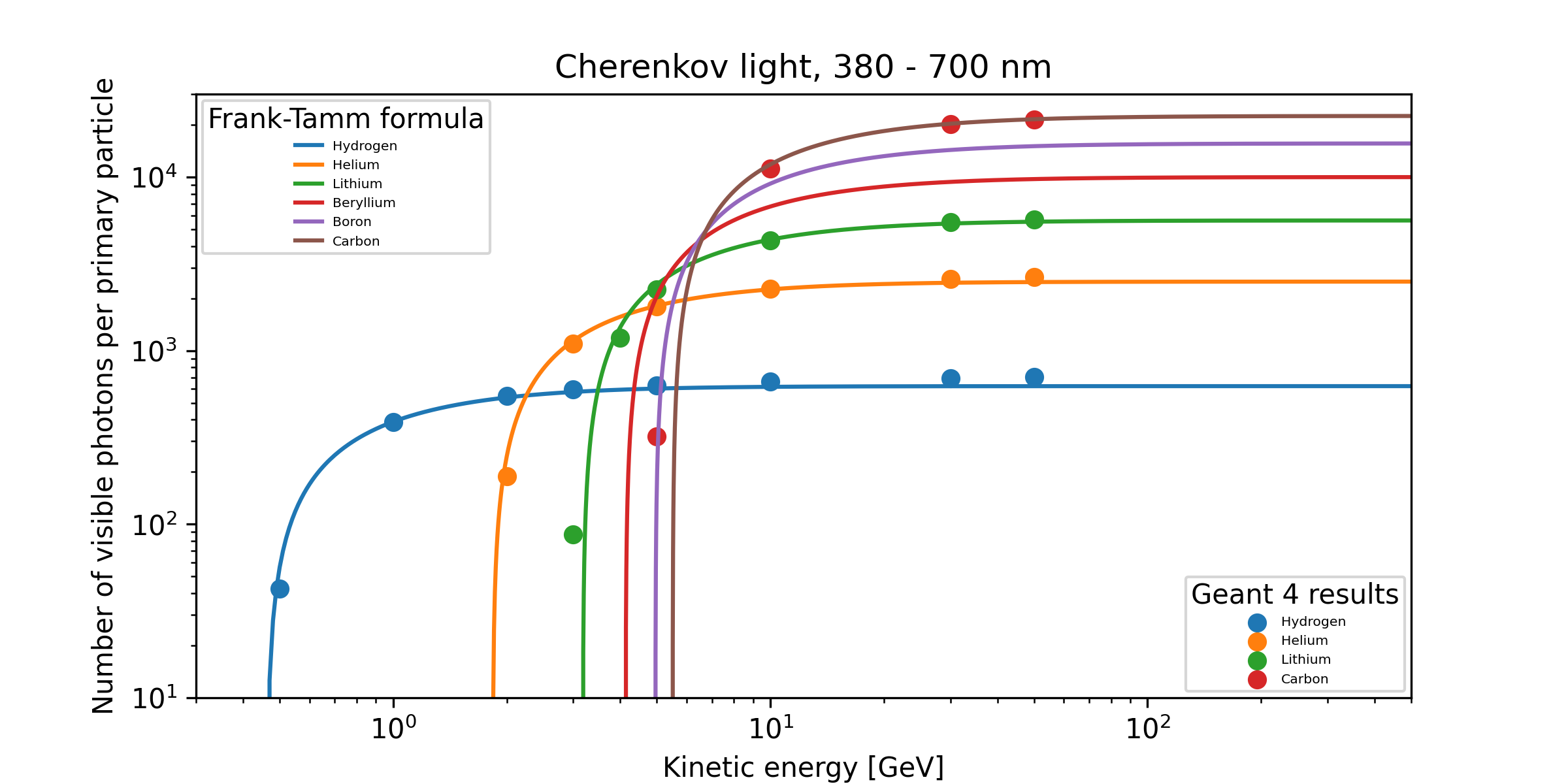}
\caption{Cherenkov light production comparison for the protons, helium, lithium, and carbon nuclei between the Geant4 simulation (circles, setup SG-Vac) and the Cherenkov light production model (lines).}
\label{fig:Cherenkov_light_simple_model_element_elements_1_6}
\end{figure}

Before going further with Cherenkovov light model we compared model results with simulation in Geant4 model. For configuration with sphere eye water model with radius 1.25 cm in world block with dimensions 2.4m x 2.4m x 2.4m and air block - 2m x 2m x 2m (SG-Vac in  table \ref{tab.2}), we compared results with Cherenkov light model for wavelength range 380 to 700 nm. Let us note, that beams of tested elements in Geant4 was not significantly scattered in air around water eye model, and cross eye model mostly along central line (z axis), thus most of elements trajectories have trajectory long 2,5 cm in water eye model. The Cherenkov light model use for this comparison trajectory in water 2,5 cm long and wavelength range between 380 to 700 nm. The protons, helium, lithium and carbon nuclei results were compared. Results are presented in the figure \ref{fig:Cherenkov_light_simple_model_element_elements_1_6}, where we can see that Geant4 mean photons intensities for selected energies  0.5, 1, 2, 3, 5, 10, 30, 50 GeV agree precisely with Cherenkov model results.  Because Geant4 values from all simulated processes agree with values from Cherenkov light production model, this additionaly confirm that there are not optical photons from other processes than from Cherenkov production. And it also confirm that we can use a Cherenkov light model for mass approximate simulations of light flashes. 

\begin{figure}[!htbp]
\noindent\includegraphics[width=\textwidth]{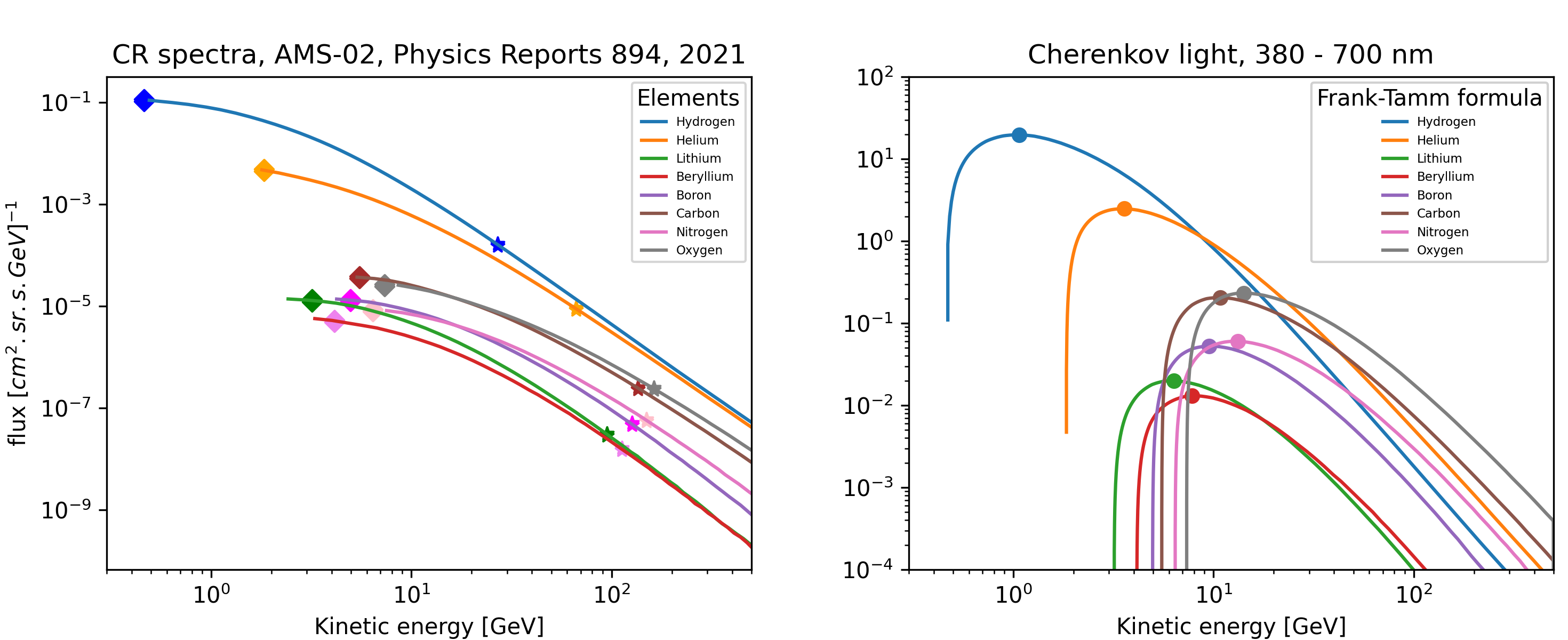}
\caption{The left panel shows cosmic ray flux spectra for nuclei from hydrogen to oxygen. Diamonds indicate the kinetic energy at which Cherenkov light production begins, and stars denote the onset of the plateau in Cherenkov light production. The right panel shows the Cherenkov light yield in the water eye model as a function of nuclear kinetic energy for elements from hydrogen to oxygen. Circles indicate the energy at which the Cherenkov light production reaches its maximum.}
\label{fig:CR_spectrum_Cherenkov_light_simple_model_element_elements_1_8_Yield}
\end{figure}

After verification of the Cherenkov light production model, the light yield was multiplied by the mean path length of particles traversing the eye, taken as 1.667 cm. This value follows from the assumption of isotropically incident cosmic rays and an eye radius of 1.25 cm. The calculation was performed using an energy step of 0.01 GeV and weighted by the cosmic-ray energy spectra.
We used AMS-02 spectra for the first eight elements of the periodic table, from hydrogen to oxygen, taken from \cite{Rb8} and spectrum of iron, the heaviest cosmic ray element characterized to date, from \cite{Rb6}. 
These AMS-02 spectra for first eight elements of the periodic table are shown in Figure \ref{fig:CR_spectrum_Cherenkov_light_simple_model_element_elements_1_8_Yield} left panel, with diamonds indicate the kinetic energy at which Cherenkov light production begins (i.e., where the
particle velocity exceeds the speed of light in water), and stars denote the onset of the plateau
in Cherenkov light production. The AMS-02 spectra were rebinned to an energy step of 0.01 GeV and folded with the Cherenkov light production evaluated with the same energy binning. This procedure yields an energy dependent yield function describing the number of Cherenkov photons produced by cosmic rays of different energies.
The resulting yield exhibits the expected behavior, with a rapid decrease beyond the maximum of photon production. This trend reflects the combination of the plateau in Cherenkov light production at high energies and the power law slope of the cosmic ray energy spectrum.

The yield functions for the first eight elements of the periodic table are shown in Figure \ref{fig:CR_spectrum_Cherenkov_light_simple_model_element_elements_1_8_Yield} right panel with the maximum of the yield functions marked by circles. Yield is shown for cosmic rays isotropicaly falling to 1 $cm^2.sr$ of eye surface per 0.01GeV bin. The figure shows that most Cherenkov photons are produced by hydrogen and helium, followed by nitrogen and carbon. 

The integral of the yield function gives the number of photons produced by cosmic rays per second, per square centimeter, and per steradian. Taking into account the eye surface area of $19.6 cm^2$, the $2\pi$ solid angle corresponding to the inward hemisphere of isotropically incident cosmic rays, the fraction of the inner eye surface covered by the retina (assumed to be 2/3), and an exposure time of one minute, we estimated the number of photons produced in the eye per minute by different cosmic ray elements that could be registered by the retina.

Table \ref{tab Cherenkov model : n of crossing,  n of photons per particle} shows number of cosmic ray crossing eye per minute (column N. of CR per minute), assuming isotropically distributed incoming cosmic ray directions (i.e. inteplanetary space at 1AU), and number of Cherenkov visible photons created per minute and number of photons per one incoming particle with mean free path 1.667 cm in eye (column N. of photons per particle). The results range from hydrogen nuclei, producing in average 317 photons per nucleus, up to oxygen nuclei, producing more than 19,000 photons per nucleus. Additional iron give 56,000 photons per nucleus.

\begin{table}
 \caption{Table presents the results of the Cherenkov light production model for cosmic ray elements from hydrogen to oxygen and iron. The maximum production energy, $T_{max}$, denotes the kinetic energy at which Cherenkov photon production is highest. Yield per minute gives the total number of Cherenkov photons produced in the eye per minute by each cosmic ray species, and N. of CR per minute gives the corresponding particle flux through the eye. N. of flashes per minute reports the predicted rate of light flashes for the two modeled visual perception thresholds, $A \ge 5$ and $Ap \ge 14$. The last column gives the mean number of Cherenkov photons produced by a single cosmic ray particle crossing the eye.}
 \label{tab Cherenkov model : n of crossing,  n of photons per particle}
 \centering
    \begin{tabular}{|c|c|c|c|c|c|}
        \hline
\textbf{Element} & \textbf{$T_{max}$} & \textbf{Yield per } & \textbf{N. of CR} & \textbf{N. of flashes} & \textbf{N. of photons}  \\ 
\textbf{number} & \textbf{[GeV]} & \textbf{minute} & \textbf{per minute} & \textbf{per minute} & \textbf{per particle}  \\ 
\textbf{ } & \textbf{ } & \textbf{ } & \textbf{ } & \textbf{$Ap\ge 5$ - $Ap \ge 14$} & \textbf{ }  \\ 
        \hline
Hydrogen & 1.07 & 305808 & 964.7 & 0.03 - 0 & 317 \\
        \hline
Helium & 3.58 & 109117 & 97.7 & 0.67 - 0 & 1117 \\
        \hline
Lithium & 6.34 & 1276 & 0.6 & 0.06 - 0 & 2293 \\
        \hline
Beryllium & 7.82 & 1107 & 0.3 & 0.10 - 0.0002  & 4098 \\
        \hline
Boron & 9.47 & 5462 & 0.8 & 0.49 - 0.015 & 6566 \\
        \hline
Carbon & 10.78 & 28266 & 2.7 & 1.84 - 0.39 & 10281 \\
        \hline
Nitrogen & 13.24 & 10236 & 0.7 & 0.51 - 0.26 & 14822 \\
        \hline
Oxygen & 14.17 & 47292 & 2.4 & 1.81 - 1.19 & 19506 \\
        \hline
Iron & 28.77 & 10451 & 0.17 & 0.13 & 56504 \\
        \hline
    \end{tabular}
\end{table}

While the number of hydrogen nuclei with energies up to 500 GeV crossing the eye per minute is relatively high (964.7 nuclei per minute), the corresponding flux decreases rapidly for heavier elements. For helium, the rate is 97.7 nuclei per minute, whereas for heavier nuclei the rates are much lower, 2.7 nuclei per minute for carbon, 2.4 for oxygen, and 0.17 for iron. 

Most of the visible light created in the eye per minute is from hydrogen and helium nuclei (column Yield per minute in table \ref{tab Cherenkov model : n of crossing,  n of photons per particle}). Overall, the table shows that a large number of light element nuclei (hydrogen and helium) cross the eye each minute, whereas only a small number of heavier nuclei, mainly carbon and oxygen, do so. The flux of heavier nuclei per minute is much closer to the event frequencies reported from Apollo missions data (Table \ref{tab.1}), the flux of lighter elements, especially hydrogen and helium, is significantly higher. This suggests that the light flashes experienced by astronauts are more likely associated with events producing several tens of thousands of photons, originating from less frequent but heavier nuclei, rather than from the much more frequent hydrogen nuclei, which typically produce only hundreds of photons per event.

To address why the light from hydrogen and helium is not perceived by astronauts, we examine the physiology of the eye in the analysis in the chapter \ref{Eye light registration-perception implementation}.

In case of heavy nuclei we have one order of magnitude more particles creating light flashes with more than 10 thousands photons per event, than Apollo numbers mentioned in table \ref{tab.1}. Which of them could create light flashes is also question of analysis with eye physiology parameters included in model, described in the next chapter \ref{Eye light registration-perception implementation}.

\subsubsection{Eye light perception implementation}
\label{Eye light registration-perception implementation}

The visual perception of light is affected not only by the number of photons reaching the retina but also by their geometry/trajectories and their absorption efficiency. The photons must activate the minimum number of rods in a specific retinal area. According to Hecht, Shlaer \& Pirenne, 1942 article \cite{Ib10} at least 5 – 14 photons must be absorbed by closely spaced rods within a small retinal area containing roughly 500 rods for creating a visual sensation. We name this condition in later text activation limit $Ap \ge 5-14$ of absorbed photons, which leads to a visual effect.

To include this activation limit in the model, we follow the intersections of the created photons trajectories in the Geant4 simulation with the retinal surface. Those intersection points were then mapped to a retina map covered with areas of $\approx$ 500 rods, which we call here $A_{500}$. A retinal integration area containing approximately 500 rods was represented by a square of 22 × 22 rods (484 rods), which differs from the target value by only 3.2 \%. A human rod cell is $\approx$ 2 microns \cite[pp.~545--546]{Rb4}  in diameter, and we approximate it by area of 0.002 mm x 0.002 mm. Thus, we check the number of photons intersecting areas $A_{500}$, every area $A_{500}$ with size 0.044 mm x 0.044 mm. 

When photons intersect the retina, we simulate the absorption process at the intersected rod by applying the quantum efficiency of rods $29\%$ from the article \cite{Rb5}. For each photon rod intersection, a uniformly distributed random number between 0 and 1 was generated. Photon absorption was accepted with probability 0.29 and rejected otherwise.


For each simulated primary particle, Geant4 provides the trajectories of all produced Cherenkov photons. The intersections of these trajectories with the retinal surface were mapped onto the $A_{500}$ retinal grid. Photon absorption was then simulated using the rod quantum efficiency. The number of absorbed photons in each $A_{500}$ area was counted and compared with the activation threshold Ap.
Because most of the photon trajectories' intersections with retina surface, and subsequent absorptions, were concentrated in central area of retinal map (most of primary particles stay close to central axis z, for simulations in setup SG-Vac), retinal surface was approximated as plane with 228 x 228 $A_{500}$ areas centered at intersection point of xy plane at z=1,25 cm at eye surface. This plane approximates $1 cm^2$ of retinal xy surface around the central axis z. We could use this approximation because most of the photons fall, and part of them are absorbed, by a few central $A_{500}$ areas. 
Averaged over 1000 simulated 30 GeV oxygen nuclei, approximately 99\% of activated $A_{500}$ areas for the activation condition $Ap \ge 5$ were located within a central retinal region of radius 0.528 mm around the central z axis. This strong localization justifies approximating the retinal surface as a plane in the central region relevant for light flash formation.

\begin{figure}[!htbp]
\noindent\includegraphics[width=\textwidth]{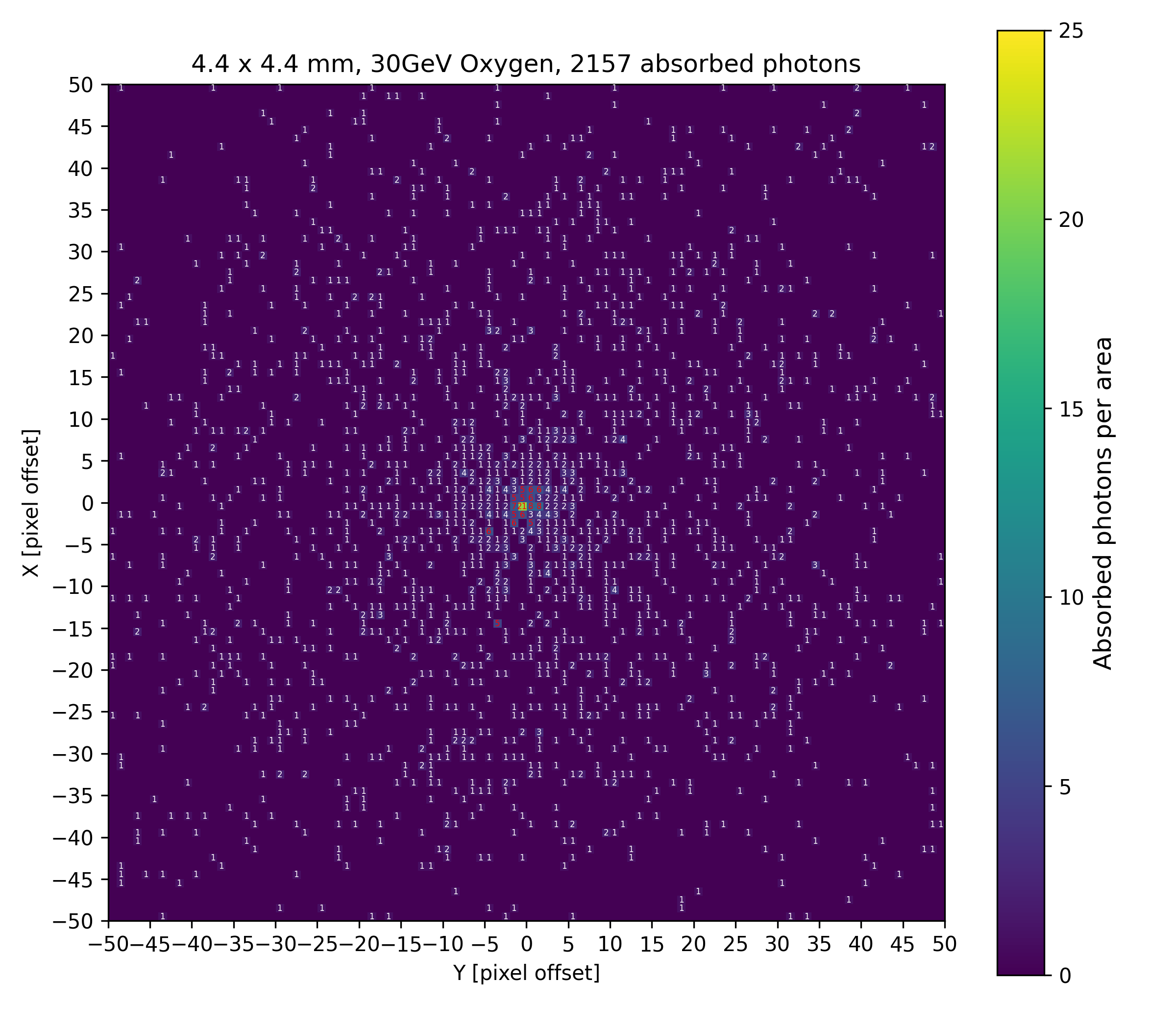}
\caption{The example for 30GeV Oxygen nucleus shows 100 x 100 $A_{500}$ areas with numbers of absorbed photons. The areas with one or more absorbed photons are marked by the number of absorbed photons (for full details, see text).}
\label{fig:Oxygen_30GeV_100x100areas_14XY_heatmap}
\end{figure}

The example for 30GeV Oxygen nucleus at figure \ref{fig:Oxygen_30GeV_100x100areas_14XY_heatmap} shows 100 x 100 $A_{500}$ areas with numbers of absorbed photons. The areas with one or more absorbed photons are marked by the number of absorbed photons. For those $A_{500}$ areas with the number of absorbed photons equal or higner than 5, the number is in red color. One could see that in a 100 x 100 $A_{500}$ areas field, what is area 4.4 x 4.4 mm, there were 2157 absorbed photons in total. Only a few in the central region, in the central 10 x 10 $A_{500}$ areas field, were activated, which we could see as areas with a red number of absorbed photons equal to or higher than 5. This is an area 0.44 x 0.44 mm, smaller more than four times than the central square millimeter. This is an area where almost all activations happen for most of the simulated particles. 

To illustrate results of applying those physiological parameters (Ap limit on $A_{500}$ areas) to results of Geant4 similations for all elements, from Hydrogen to Oxygen and Iron, for selected primary nucleus kinetic energies including  1, 2, 3, 5, 10, 30 and 50 GeV, and those additional needed to better catch shape of activations dependency on energy, in model, we show examples of results for selected kinetic energies 10 and 30GeV.

The analysis shows a complex picture, where the number of activated areas at the retinal surface (retinal areas exceeding the activation threshold) vary with energy and primary particle type. While for light elements there were rarely some activations, in the case of heavier elements, there were activations in almost every simulated case. 

\begin{figure}[!htbp]
\noindent\includegraphics[width=\textwidth]{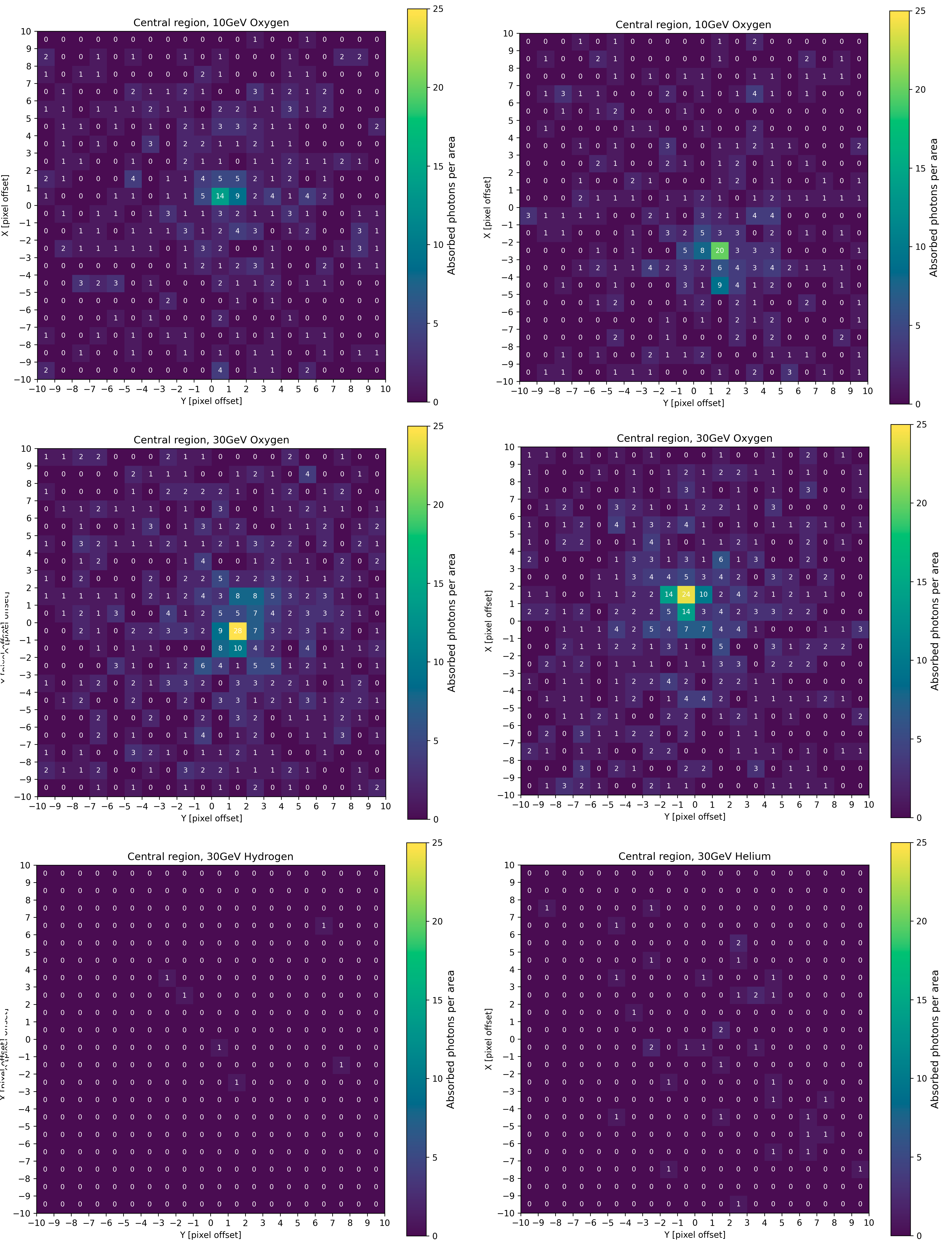}
\caption{Number of absorbed photons for 10GeV and 30 GeV oxygen and 30GeV hydrogen and helium, in the central 20x20 $A_{500}$ areas covers approximately a central square millimeter, i.e., 0.88 x 0.88 mm (for full details, see text).}
\label{fig:Hydrogen_Helium_10GeV_Oxygen_10_and_30GeV_examples_of_areas_absorption_counts}
\end{figure}

In the figure \ref{fig:Hydrogen_Helium_10GeV_Oxygen_10_and_30GeV_examples_of_areas_absorption_counts} we  present examples of central 20x20 $A_{500}$ areas (approximately central square millimeter, i.e. 0.88 x 0.88 mm) for 10GeV and 30GeV Oxygen and 30GeV hydrogen and helium, with number of absorbed photons.

In the upper two figures are shown two examples of absorbed photons from Cherenkov photons produced by 10GeV Oxygen nucleus. In both upper panels, we see cases when for $Ap \ge 14$ just one area will be activated, and for $Ap \ge 5$ five areas will be activated on the left figure, and 6 on the ritht one.
For 10GeV Oxygen is the number of created photons along approximately a 2.5cm long trajectory in average 11028 photons. This shows that only a small part of the produced photons is perseived as light flash. 71 percent is not absorbed due to rods quantum efficiency, and from those 29 percent absorbed, most of them end as one photon absorbed at some $A_{500}$ area. Thus, in both cases for 10GeV Oxygen, depending on the Ap limit, only a few areas are activated. For $Ap \ge 5$, 38 photons was seen on the left panel and 53 photons on the right panel. In both presented examples, approximately 98.5 percent of absorbed photons does not end as visual perception. For $Ap \ge 14$ approximately 99.5 percent of absorbed photons is not perseived.

Two examples of 30GeV Oxygen Cherenkov absorbed photons are presented in the middle row of \ref{fig:Hydrogen_Helium_10GeV_Oxygen_10_and_30GeV_examples_of_areas_absorption_counts}. On the left panel, 121 photons (15 activated areas) and on the right panel, 97 photons (11 activated areas) were absorbed in activated $A_{500}$ areas for $Ap \ge 5$. Because, for 30GeV Oxygen on average, 33381 photons are created along a 2.5 cm path in the eye water model, 99 percent of absorbed photons were not visually perceived. In the case of $Ap \ge 14$, there was only one area activated on the left panel and three at the right panel. The number of perceived photons decreases to approximately 0.2-0.4 percent.

And in the bottom row, we present examples of absorbed photons from 30GeV hydrogen and helium nucleus. There are no activated $A_{500}$ areas for the hydrogen case in the left panel. Few areas absorb one photon. For helium, on the right panel, we see simmilar situation, now with a few $A_{500}$ areas absorbed two photons, but again no activated areas. The situation is different from the situation for the oxygen nucleus, there is no visual perception in the hydrogen and helium cases.

\begin{figure}[!htbp]
\noindent\includegraphics[width=\textwidth]{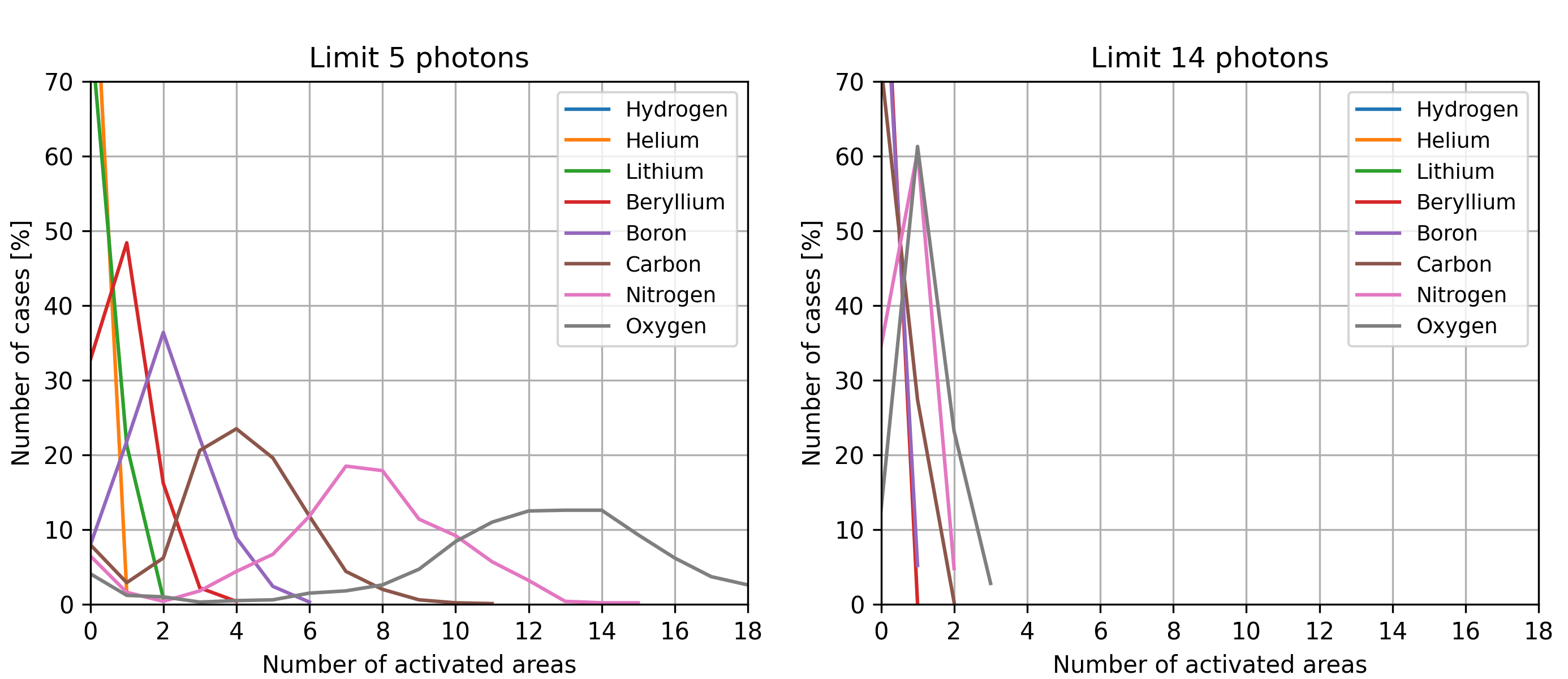}
\caption{Normalized distributions of the number of activated $A_{500}$
areas produced by 30 GeV cosmic ray nuclei from hydrogen to oxygen. The left panel corresponds to the visual perception threshold $Ap \ge 5$, and the right panel to $Ap \ge 14$.}
\label{fig:number_of_activated_areas_histogram_30GeV_limit_5_and_14}
\end{figure}

To illustrate the number of activated $A_{500}$ areas dependency on Ap limit and element kinetic energy, we present a case for 30 and 10GeV.
The figure \ref{fig:number_of_activated_areas_histogram_30GeV_limit_5_and_14} shows the number of activated areas for 30GeV elements, from hydrogen to oxygen. The hydrogen simulations use 10 thousand primary particles, for other elements, one thousand particles passed through the eye model were used (setup x from table y). As is shown in figure \ref{fig:number_of_activated_areas_histogram_30GeV_limit_5_and_14} for the case with a lower activation limit $Ap \ge 5$ photons, there is zero activations from 10 thousand hydrogen 30GeV particles. There is no case when more than 5 photons were absorbed by any of the $A_{500}$ retinal areas. But for Oxygen in most cases there were 10-15 areas activated (in 66 percent cases) and for Nitrogen in most cases 6-10 areas were activated (69 percent). For Carbon, there were, in most cases 3-6 (76 percent), and for Boron 1-4 activations (89 percent). In most cases, we mean here those around the maximum of the element histogram, summing histogram bins with values over 9 percent.

When the activation limit $Ap \ge 14$ photons for activations of $A_{500}$ areas was applied, the situation change dramatically/significantly. There were no activations for 30GeV hydrogen, helium, and lithium. For Beryllium, there was just one case of 30GeV nucleus from one thousand, whith activation of one area. For 30GeV Boron, there was only one area activation appearing in 5 percent of cases. For Carbon, there was only one (27 percent) and two area activations (0.3 percent), for Nitrogen, only one area (61 percent) and two area activations (5 percent). Only for Oxygen, there were a few cases when three areas were activated (3 percent). The one in 61 percent of cases and two in 23 percent of simulated Oxygen cases.

\begin{figure}[!htbp]
\noindent\includegraphics[width=\textwidth]{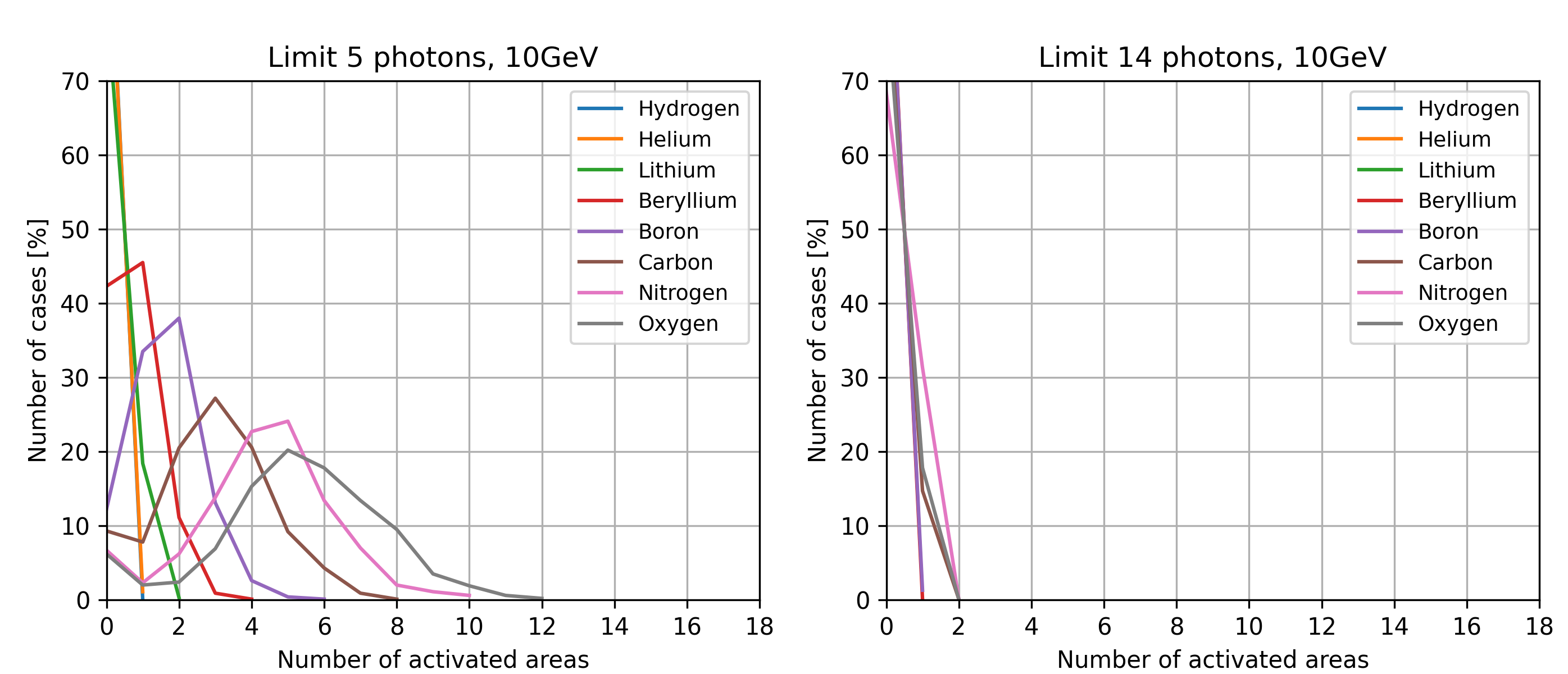}
\caption{Normalized distributions of the number of activated $A_{500}$
areas produced by 10 GeV cosmic ray nuclei from hydrogen to oxygen. The left panel corresponds to the visual perception threshold $Ap \ge 5$, and the right panel to $Ap \ge 14$.}
\label{fig:number_of_activated_areas_histogram_10GeV_limit_5_and_14}
\end{figure}

For 10 GeV particles and a limit $Ap \ge 5$ photons for activation, the situation is shown in figure \ref{fig:number_of_activated_areas_histogram_10GeV_limit_5_and_14} left panel. One could notice that for 10 GeV Oxygen, the number of activated areas changed from 10- 15 areas in 66 percent of primaries for 30 GeV nuclei to 4-8 areas activated in 76 percent of cases. For 10 GeV Nitrogen, the change from 6-10 areas activated in 69 percent of primaries for 30 GeV nuclei to 74 percent activations with 3-6 areas. And for 10 GeV Carbon, it decreases from 30 GeV case to 2-5 activations in 78 percent of cases.

For 10 GeV particles and a limit $Ap \ge 14$ photons for activation, there are at maximum two areas activated in case of Oxygen (0.7 percent), Nitrogen (0.1 percent), and Carbon (0.1 percent). One area was activated for Oxygen in 43 percent of cases, for Nitrogen in 31 percent, and for Carbon in 15 percent of cases.

For every element, we evaluated activations probability curve $P$ describing the probability of activation of areas with more or equal than 5 or 14 absorbed photons, which is the probability of light event visual sensation. The simulated activation probabilities were linearly interpolated onto a 0.01 GeV energy grid. The resulting probability of $A_{500}$ areas activation for the i-th element $P_i (T_{kin}, Ap)$ with activation condition with number of absorbed photons Ap higher or equal to 5 or 14, is a function of kinetic energy and the limit for number of absorbed photons Ap. 

\begin{figure}[!htbp]
\noindent\includegraphics[width=\textwidth]{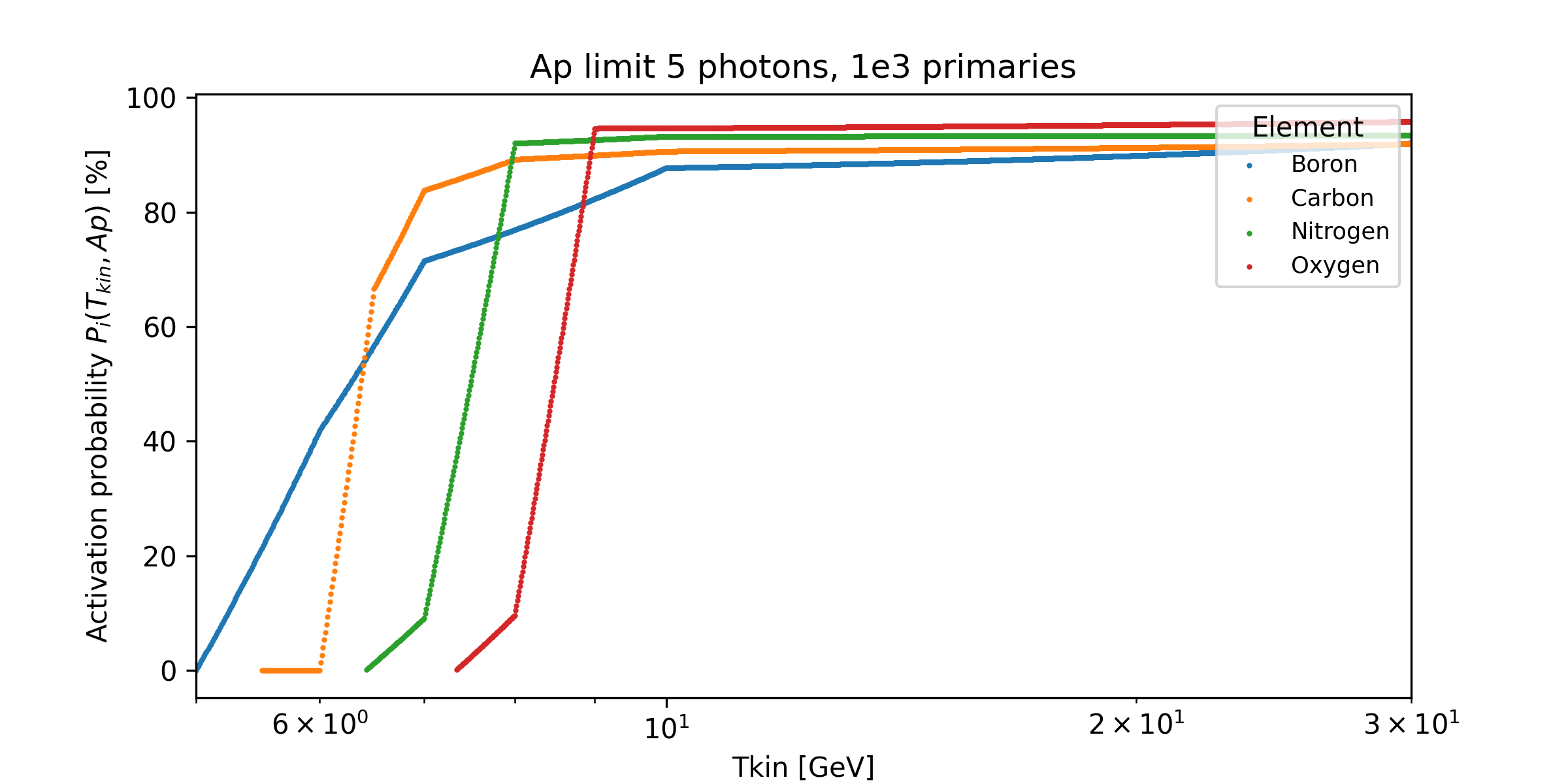}
\caption{Probability of $A_{500}$ area activation, $P_i (T_{kin}, Ap)$, as a function of kinetic energy for boron, carbon, nitrogen, and oxygen nuclei, assuming the visual perception threshold $Ap \ge 5$.}
\label{fig:activation_probability_curves_limit_5}
\end{figure}

The example of $P_i (T_{kin}, Ap)$ for boron, carbon, nitrogen, and oxygen and $Ap \ge 5$ is shown in figure \ref{fig:activation_probability_curves_limit_5}. The probability curve starts at the limit of kinetic energy where the nucleus starts producing Cherenkov light, rapidly rises to values close to a plateau around 80-90 percent. 

Probability $P_i (T_{kin}, Ap)$ was used to evaluate number of light flashes seen by eye. In model is number of light flashes $N_{LF,i}$ for element $i$ evaluated as

\begin{equation}
N_{LF,i, Pn} = G.\int_{0GeV}^{500GeV}  I_{CR,i}(T_{kin}) . P_i(T_{kin}, Ap)\, dT_{kin} 
\end{equation}

where $I_{CR,i}(T_{kin})$ is cosmic rays differential intensity of element $i$,  and G is geometry constant. The geometry constant G includes the surface of the eye (19,6 $cm^2$), part of the eye surface sensitive to light (2/3), and the solid angle 2$\pi$ from which cosmic rays approach the eye surface.

Results are summarized in \ref{tab Cherenkov model : n of crossing,  n of photons per particle} in column \textit{N. of flashes per minute}. 
For $Ap >=5 $, the number of light flashes per minute from all elements from Hydrogen to Oxygen is 5.51, for $Ap >= 14$, it is 1.86.

The physiology of light event registration by the human eye with an Ap limit range of 5 to 14 absorbed photons lead to quite a different situation for light flashes production. However, some clear conclusions remain the same for the whole Ap range. We notice that ligh from light elements are effectively suppressed and not visible to astronauts due to the geometry of produced Cherenkov light and effectivity of photons absorption. This lead to the answer to the question why we do not see light flashes at Earth's surface explained further in chapter \ref{Muon induced light flashes at the Earth's surface}. The final result of whole model parts put together shows that most light flashes seen in space are created by Carbon and Oxygen.

The observed characteristics of astronaut light flashes, of which approximately 66\% are reported as spot or starlike, are also consistent with Cherenkov light production in the eye.

Another types of light flashes are streak (25\%) and cloud (8\%). While streak could be case when particle crossing eye in parrallel with local retina surface, cloud could be created by multiparticle interaction with eye in perceiving time window or by heavier elements producing more light than Carbon and Oxygen. We address both possibilities in next two subchapters.

This could also mean that 10 thousands particles is still not enough to create light flash experience. The number of those with 50 thousand photons per event is comparable with Apollo numbers. There are not enough particles creating higher numbers of photons by Cherenkovov radiations to reproduce Apollo results. If light flash experience need few hundred thousand or million or more photons, than light flashes are not Cherenkovov light effect. 

The final aspect considered is the reported color of astronaut light flashes. According to the survey of \cite{Ib4}, white flashes are by far the most frequently reported (41 responses), whereas all other colors combined account for only 14 responses. If blue flashes, which are also consistent with the broadband Cherenkov spectrum, are included, the ratio becomes 44 white or blue responses versus only 11 responses of other colors. In our simulations, Cherenkov photons are produced over the entire visible wavelength range (380–700 nm), with a continuous spectrum that is more intense at shorter wavelengths, consistent with the expected Cherenkov emission spectrum. Such broadband emission is therefore expected to be perceived predominantly as white or bluish-white. The relatively small number of reports of yellow, orange, green, or red flashes may reflect the limitations of human color perception under dark adapted conditions or indicate contributions from additional radiation induced mechanisms. Thus, while the simulated Cherenkov spectrum is consistent with the predominance of white and blue light flashes reported by astronauts, it does not necessarily account for every observed color.

\subsubsection{Multiple particles crossing eye in one 0.1 sec time window}

The filtration effect of the Ap limit effectively makes visible light created by the hydrogen and partly also by helium nucleus not perceived. Nevertheless, because the number of these light elements crossing eye is high (column \textit{N. of CR per minute} in table \ref{tab Cherenkov model : n of crossing,  n of photons per particle}), we disscuss in this subchapter possibility to perceive light from multiple particles ligh flashes in 0.1 scond long perceiving time window.

Two or more particles could be perceived as a single light flash if they arrive within the eye temporal integration window. With 964.7 hydrogen nuclei per minute, there are multiple crossings per time window assumed to be 0.1 second. Assuming Poisson process, there are multiparticle crossing per 0.1 second window, with number of particles crossing eye per minute shown in Table \ref{tab Cherenkov model : n of crossing,  n of photons per particle} for hydrogen, helium and carbon. For example, for protons, there is 287 multiparticle crossing per minute (probability $P(\ge 2)$, while there is almost every second integration window with 4 or more crossing protons $P(\ge 4)$. Seven or more protons (hydrogen nuclei) during integration time window we could see approximately once per minute $P(\ge 7)$, which give $\approx 2300$ photons created per 0.1 second temporal integration window. However, chance of overcome Ap limit for hydrogen was very low, generally much less than 1 particle from one thousand of primary particles create light flash sensation even for $Ap \ge 5$. The isotropically incoming cosmic rays, fo case of 7 particles in 0.1 second time window will unlikely overlap their most enlighted areas.
There are seven multiparticle crossing of helium per minute, but only one multiparticle crossing from Carbon per almost three hours. This show that multiparticle light flashes from heavier than helium elements are very rare, the astronauts see from heavy elements only light flashes created by one nucleus per perceiving time window. The Table \ref{tab simple model : Multiparticle LF} summarizes the multiparticle light flashes per minute results with assumed Poisson process for hydrogem helium and carbon.

\begin{table}
 \caption{Table showing the frequency of multiparticle light flashes per minute for hydrogen, helium, and carbon nuclei. Values smaller than one multiparticle light flash per 10 years are indicated by a dash. Columns labeled “A” correspond to frequencies in interplanetary space at 1 AU. Columns labeled “B” correspond to low Earth orbit in high geomagnetic latitude regions with low cut-off rigidity, where the particle spectrum is minimally affected by the geomagnetic field, but the total intensity is reduced by approximately a factor of two relative to interplanetary space due to the Earth’s shadow.}
 \label{tab simple model : Multiparticle LF}
 \centering
    \begin{tabular}{|c|c|c|c|c|c|c|}
        \hline
\textbf{N. of nuclei} & \textbf{$Hydrogen_{A}$} & \textbf{$Helium_{A}$} & \textbf{$Carbon_{A}$} & \textbf{$Hydrogen_{B}$} & \textbf{$Helium_{B}$} & \textbf{$Carbon_{B}$}\\ 
\textbf{per 0.1 sec} & \textbf{[$min^{-1}$]} & \textbf{[$min^{-1}$]} & \textbf{[$min^{-1}$]} & \textbf{[$min^{-1}$]} & \textbf{[$min^{-1}$]} & \textbf{[$min^{-1}$]}\\ 
        \hline
$P(>=1)$ & 479.8 & 90.16 & 2.694 & 331.5 & 46.91 & 1.348\\
        \hline
$P(>=2)$ & 286.6 & 7.141 & 0.00606 & 115.6 & 1.884 & 0.00152 \\
        \hline
$P(>=3)$ & 131.2 & 0.382 & 9.08e-06 & 28.79 & 0.0508 & -\\
        \hline
$P(>=4)$ & 47.94 & 0.0154 &  - & 5.539 & 0.00103 & - \\
        \hline
$P(>=5)$ & 14.47 & 0.0005 &  - & 0.865 & 1.67e-05 & - \\
        \hline
$P(>=6)$ & 3.708 & 1.35e-05 &  - & - & - & - \\
        \hline
$P(>=7)$ & 0.824 &  -  &  - & - & - & - \\
        \hline
$P(>=8)$ & 0.161 &  -  &  - & - & - & - \\
        \hline
$P(>=9)$ & 0.0283 &  -  &  - & - & - & - \\
        \hline
$P(>=10)$ & 0.00447 &  -  &  - & - & - & - \\
        \hline
    \end{tabular}
\end{table}

Results from table \ref{tab simple model : Multiparticle LF} suggests that multiparticle crossing of hydrogen nuclei is unlikely to create the light flashes. The heavier elements almost do not have multiparticle crossings in the perceived time window. However, helium is a specific case, when three times per minute there are three helium nuclei crossing the eye in the perceiving time window (in interstellar space, at the low Earth orbit, it is half, i.e., 1.5). Those could be connected to some multidots light flashes.

\subsubsection{Elements heavier than Oxygen}

Among nuclei heavier than oxygen, only iron has a sufficiently high flux to contribute significantly to the Apollo light-flash frequency ($\approx$ 0.2 LF/min; see table \ref{tab.1}). Based on the AMS-02 spectrum, iron contributes approximately 0.17 retinal crossings per minute. Heavier nuclei are approximately an order of magnitude less abundant and therefore are expected to provide only a minor contribution to the total light flash rate.

Examples of retinal activation patterns produced by 40 and 50 GeV iron nuclei are shown in Figure 15. For the 40 GeV iron nucleus and threshold $Ap \ge 5$, 588 $A_{500}$ areas were activated, with 5135 absorbed photons (left panel). For the 50 GeV case with the same threshold, 954 $A_{500}$ areas were activated with 8019 absorbed photons (right panel).

To characterize the distribution of activated retinal regions, we simulated 1000 iron nuclei at energies of 40 and 50 GeV. The resulting distributions of activated $A_{500}$ areas for thresholds $Ap \ge 5$ and $Ap \ge 14$ are shown in Figure \ref{fig:number_of_activated_areas_histogram_40-50GeV_Iron_limit_5_and_14}.
For $Ap \ge 5$, the distribution peaks at approximately 600 activated $A_{500}$ areas for 40 GeV nuclei and approximately 900 activated areas for 50 GeV nuclei. Increasing the threshold to $Ap \ge 14$ reduces the number of activated regions substantially. The distributions peak near 60 activated areas for 40 GeV and 90 activated areas for 50 GeV.

With those numbers of activated areas $A_{500}$, in hundreds for lower AP limit, and over 50 for Ap limit 14, the Iron nuclei with energy over 40GeV effectively produces the cloud like light flash visual effect. 

Even at the higher activation threshold, iron nuclei above 40 GeV activate tens of retinal regions simultaneously. At the lower threshold, activation extends over several hundred $A_{500}$ areas, suggesting a spatially extended (“cloud-like”) visual perception rather than a localized flash.

\begin{figure}[!htbp]
\noindent\includegraphics[width=\textwidth]{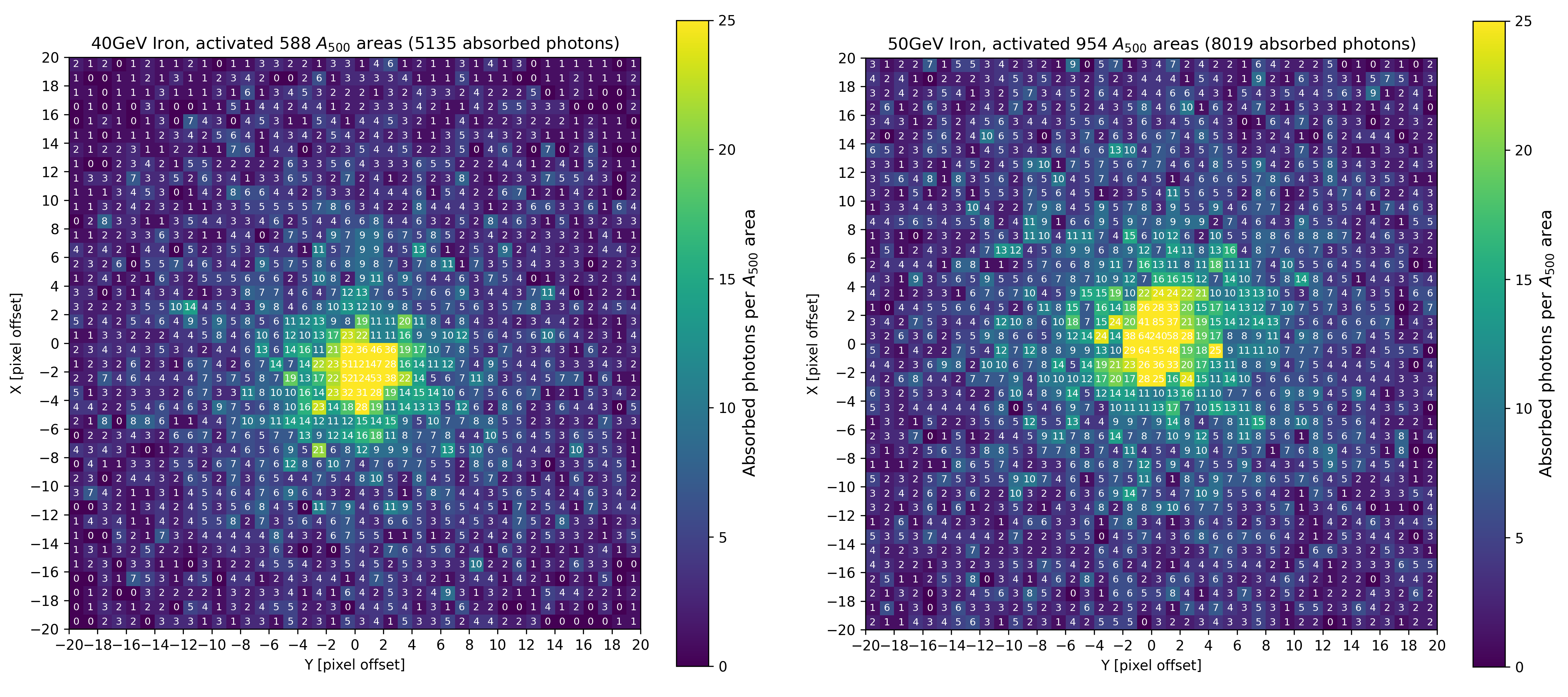}
\caption{Examples of retinal activation patterns produced by 40 GeV (left) and 50 GeV (right) iron nuclei. Colors indicate the number of absorbed photons in each $A_{500}$ area.}
\label{fig:40_50GeV_Iron_examples}
\end{figure}

\begin{figure}[!htbp]
\noindent\includegraphics[width=\textwidth]{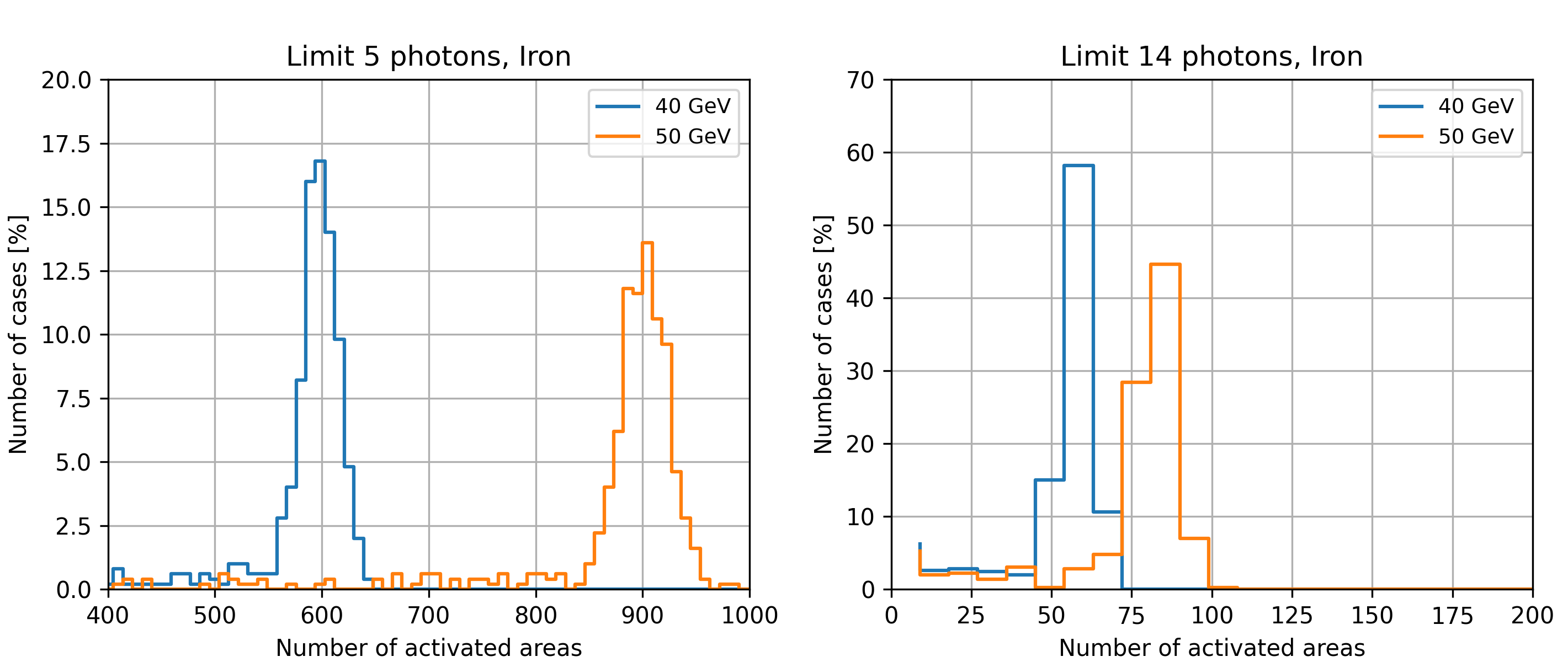}
\caption{Distributions of the number of activated $A_{500}$ areas for iron nuclei with energies of 40 and 50 GeV. Left panel: $Ap \ge 5$. Right panel: $Ap \ge 14$.
}
\label{fig:number_of_activated_areas_histogram_40-50GeV_Iron_limit_5_and_14}
\end{figure}


Using the iron energy spectrum measured by AMS-02 \cite{Rb6}, we estimated the light flash occurrence rate produced by iron nuclei in model described in subchapter \ref{Eye light registration-perception implementation}. For the threshold $Ap \ge 5$, the predicted rate is 0.17 $LF/min$. Restricting the spectrum to energies above 40 GeV yields 0.16 $LF/ min$, while energies above 50 GeV contribute 0.14 $LF/min$.

These results indicate that most iron induced light flashes originate from nuclei with energies above 40 GeV. The predicted number of iron light flashes per minute is very similar for the $Ap \ge 5$ and $Ap \ge 14$ thresholds. In the case of iron, the change in the threshold primarily affects the total number of absorbed photons and the extent of retinal activation.

Interestingly, this result may also provide an alternative interpretation of the observations by \cite{Ib4}, who suggested that approximately 5–10\% of astronaut light flashes originate from Cherenkov radiation based on the occurrence of diffuse "cloud" or "blob" morphologies. In our simulations, iron nuclei contribute a similar fraction of the total modeled LF rate while producing the largest activated retinal areas. This raises the possibility that the reported diffuse light flashes correspond predominantly to Cherenkov events induced by heavy nuclei such as iron, whereas Cherenkov light produced by lighter nuclei results in smaller retinal activation patterns and different perceived morphologies. Although this interpretation remains speculative, it offers a physically motivated explanation of the observed flash morphology.

\subsubsection{Heliosphere, magnetosphere and Earth shadow influence}

The previous sections evaluated light flash production assuming the interplanetary cosmic ray spectra measured by AMS-02 experiment. Astronauts in low Earth orbit experience a modified radiation environment because Earth occultation, and geomagnetic shielding. In this section we evaluate how these effects modify the predicted light flash rate.

The previous results were evaluated for AMS-02 spectra \cite{Rb6,Rb8}, which describe cosmic ray intensities in interplanetary space at Earth's orbit (1 AU from the Sun). The spectra were measured between May 2011 and May 2018 and represent a seven year average.

First, the cosmic ray spectrum varies over the solar cycle. The highest cosmic ray intensities are recorded during solar minimum periods. The change in comparison with the used AMS-02 proton averaged spectrum is approximately $\pm 1/4$ in total flux from 1 to 100 GV around the averaged spectrum. For proton flux from approximately 1500 protons per $[m^2.s.sr.GV]$ in 2014 to $\approx 2800$ protons per $[m^2.s.sr.GV]$ in 2019 \cite{Rb3}. For Helium flux from approximately 180 nuclei per $[m^2.s.sr.GV]$ in 2014 to $\approx 280$ helium nuclei per $[m^2.s.sr.GV]$ in 2019 \cite{Rb10}. This will lead to approximate changes in the number of evaluated light flashes by a similar factor $\approx 1/4$. For heavier elements, there are only averaged spectra published for longer periods, not allowing estimation of variability during the solar cycle.

Another important effect is Earth occultation.
While cosmic rays in interplanetary space at 1 AU are almost isotropic, and could be, for light flashes estimation purposes, approximated as isotropic, the flux in the Earth's magnetosphere is not isotropic. In Earth's low orbit, particles arrive predominantly from the upward hemisphere, but not from the nadir hemisphere. Trajectories originating below the local horizon (from nadir and other directions crosing low orbit from down) are blocked by Earth, thus intensities at the low orbit at high geomagnetic latitudes are approximately half of the intensity in the interplanetary space. 

And importantly, there is an effect on the number of observed Light flashes from Earth's magnetosphere shielding, when astronauts observe light flashes inside the Earth's magnetosphere, usually from low orbit. The intensity of cosmic rays at Earth's orbit depends on the transparency of the magnetosphere, which is different for different positions at the Earth's surface or in Earth's orbit. Transparency depends primarily on geomagnetic latitude, at high geomagnetic latitudes, where cutoff rigidities are low, cosmic rays freely reach Earth's orbit or the top of the atmosphere. Some regions under the International Space Station have low cutoff rigidities. Specifically, regions in North America near the 51 degrees of geodetic latitude, where in the longitudinal region between approximately 245 and 300 degrees, the vertical effective cutoff rigidity is lower than 1GV \cite{Rb9}.

\begin{figure}[!htbp]
\noindent\includegraphics[width=\textwidth]{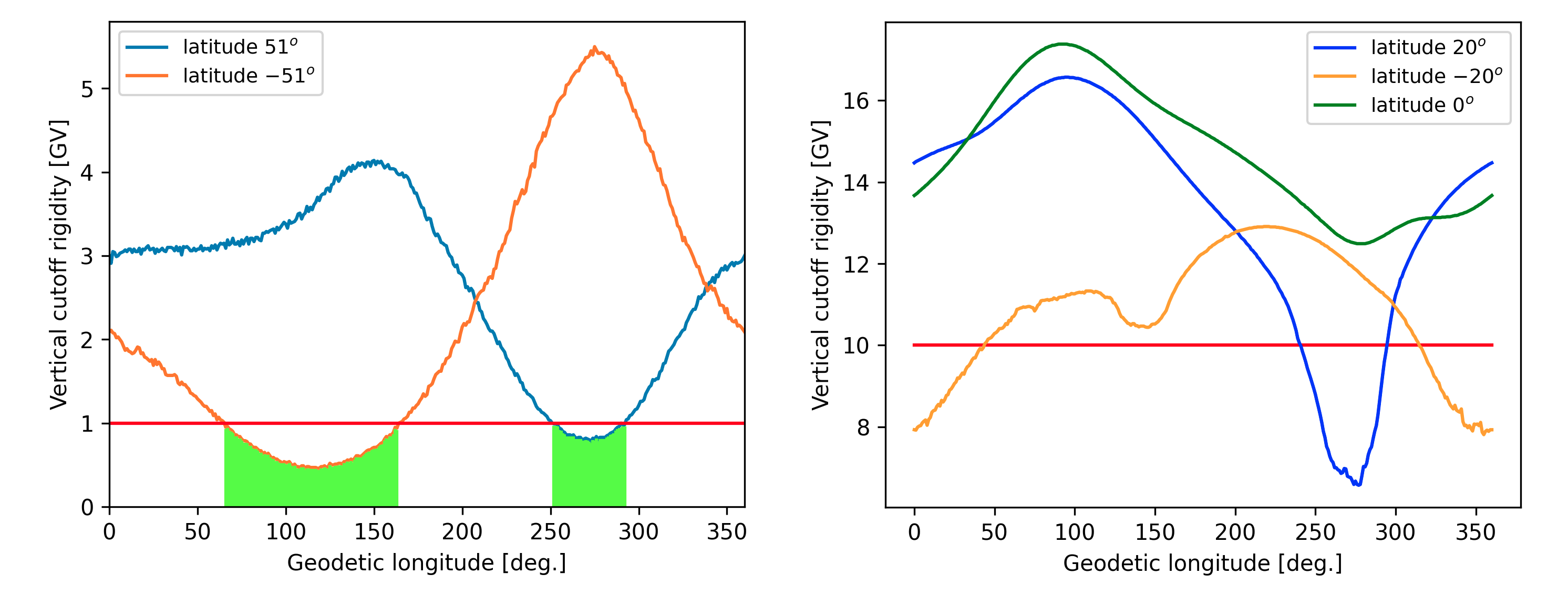}
\caption{Vertical cut-off rigidities for protons obtained from COR system simulations \cite{Rb9}, evaluated using the internal geomagnetic field model IGRF for 1 January 2010 at 00:00 UTC \cite{Rb11}, are shown for selected latitudes as a function of geodetic longitude in the upper panels. The left panel presents cut-off rigidities at latitudes +51 degrees  and -51 degrees, corresponding to the northernmost and southernmost positions of the International Space Station orbit. The right panel shows cut-off rigidities in the equatorial region for latitudes -20 deg., 0 deg., and +20 deg.. The green shaded areas in the upper-left panel indicate longitude ranges where the vertical cutoff rigidity is below 1 GV, corresponding to a proton kinetic energy of approximately 0.433 GeV.}
\label{fig:vertical_cutoff_rigidity_lat_51_m51_20_m20_0_ver_2}
\end{figure}

In equatorial regions, the proton flux is reduced by the geomagnetic field shielding, whereas in regions of low geomagnetic latitude under the ISS orbit, the conditions are closer to those in interplanetary space. Consequently, in equatorial regions, light flashes produced by protons are expected to be less frequent, while in the low geomagnetic latitude region over North America and the Indian Ocean, they should be present more often. The dependence of vertical cut-off rigidities on geodetic longitude for northern and southern latitudes of Earth low orbit is shown in Figure \ref{fig:vertical_cutoff_rigidity_lat_51_m51_20_m20_0_ver_2}. The green shaded areas  indicate longitude ranges where the vertical cut-off rigidity is below 1GV. 

\begin{figure}[!htbp]
\noindent\includegraphics[width=\textwidth]{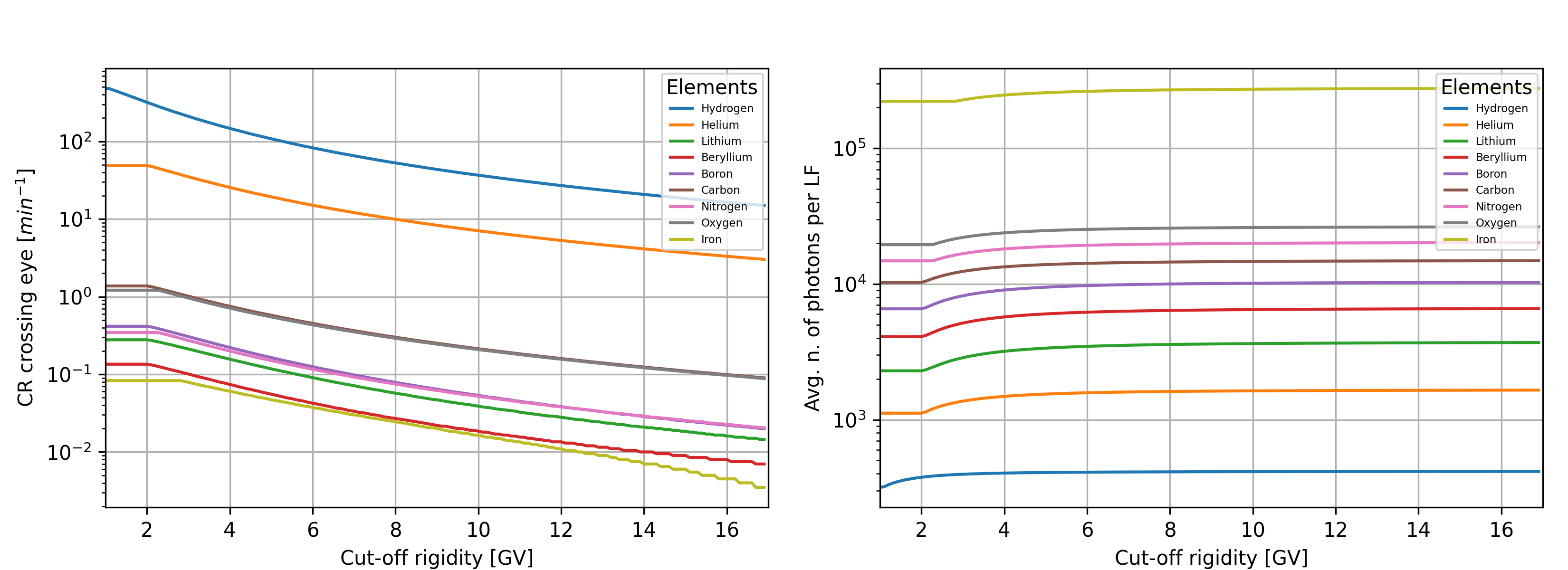}
\caption{Cosmic ray crossing eye at different cut-off rigidities (left panel) for elements from Hydrogen to Oxygen and Iron. The right panel shows the average photons produced per particle crossing the eye as a function of cut-off rigidity for the used elements.}
\label{fig:CR_and_Avg_n_of_hotons_per_LF__vs_geomagnetict_cutoff}
\end{figure}

Using a model of light flash production described in the previous text with cosmic ray spectra of cosmic rays above setted cut-off rigidities, we could show the dependency of the number of light flashes (LF rate) on the cut-off rigidity. Figure \ref{fig:CR_and_Avg_n_of_hotons_per_LF__vs_geomagnetict_cutoff} shows how number of cosmic ray crossing the eye per minute change with cut-off rigidity for elements from Hydrogen to Oxygen and Iron at low Earth orbit. Note that in low cut-off regions with small cut-off rigidities for Iron is is 0.083 nuclei crossing eye per minute (half of the number in interstellar space). This number decreases approximately ten times (to 0.007 eye crossing per minute) in regions with high cut-off rigidity over 14GV.

\begin{figure}[!htbp]
\noindent\includegraphics[width=\textwidth]{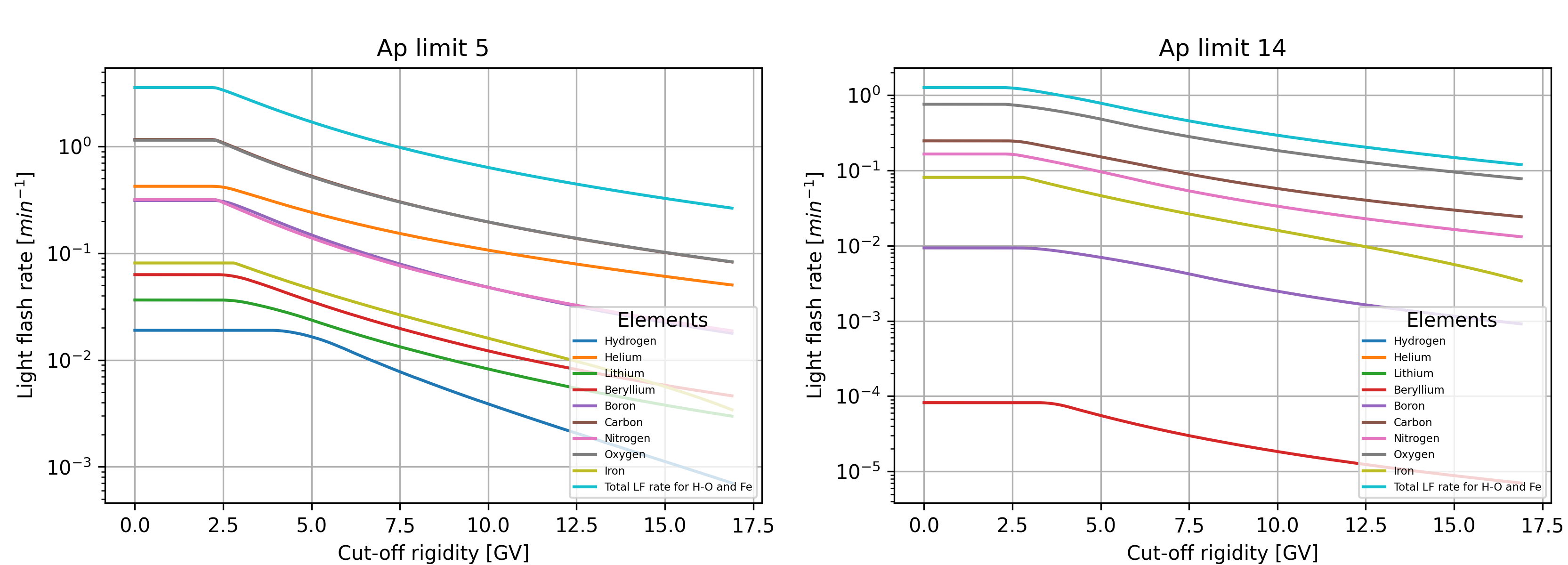}
\caption{The number of light flashes at low Earth orbit for different cut-off rigidities. Left panel: $Ap \ge 5$. Right panel: $Ap \ge 14$.
}
\label{fig:LF_vs_geomagnetic_cutoff_limit_5_and_14}
\end{figure}

The LF rate per minute at low Earth orbit at different geomagnetic rigidities is shown in figure \ref{fig:LF_vs_geomagnetic_cutoff_limit_5_and_14} for threshold $Ap \ge 5$ in the left panel and for limit $Ap \ge 14$ in the right panel. The LF rate for Oxygen and Carbon for the lower threshold $Ap \ge 5$ changes from  1.15 and 1.17 per minute in low cut-off rigidity regions to 0.11 LF per minute in high rigidity regions with cut-off rigidity over 14GV. The number of Iron LF per minute changes from 0.081 to 0.007 LF per minute. The change in LF per minute frequency, approximately 10 times reduction,  between low and high cut-off rigidity regions for the lower threshold stay similar also for the higher threshold. For $AP \ge 14$ for Carbon from 0.25 LF per minute to 0.03 LM/min. between the low and high cut-off rigidity region, and for Oxygen from 0.75 LF per minute to 0.11 LF/min. Noticeably, for Iron in the higher AP threshold case, the intensity of LF stay same as for the lower threshold, changing from 0.081 to 0.007 LF per minute between low and highg rigidity regions. The interesting result for Iron, where LF frequency does not depend on AP limit, allow us to propose experimental verification of this result. According to the model output, the number of most bright flashes from Iron nuclei, which are likely clasify by astronauts as cloud-like, changes between low and high rigidity regions 10 times. From one LF in approximately 12 minutes in a region with low cut-off rigidity under 2GV, to one LF in 143 minutes in a region with high cut-off rigidity over 14 GV.
The experiment could be completed with observations over the Indian Ocean (approximately at longitudes from 50 to 150 degrees, in latitudes between 0-20 degrees) with high cut-off rigidities, followed by observations over North America, with low cut-off rigidities. 

If no difference between low and high cut-off regions is observed, the present Cherenkov model would be challenged.

The result of approximately 10 times reduction of the LF rate between low and high cut-off rigidity is consistent with the results of the Sileye-2 experiment published in the upper panel of Figure 1 from \cite{Ib5} (green squares in it). 


However, results for the South Atlantic anomaly region (SAA hereafter), where the Sileye-2 experiment shows higher light flash frequency than in other regions shown in Figure 1 of \cite{Ib5}, is not reproduced by the present Cherenkov model. Using trapped proton spectra above 70MeV measured by the PAMELA experiment published in \cite{Rb12} in the model of LF production leads to a negligible LF frequency of less than one LF per weeks. Other nuclei have negligible intensities in SAA. As a conclusion about SAA light flashes, we could say that light flashes observed in the SAA region are caused by another mechanism, not by Cherenkov light production.

Another suggested experiment should be done in high geomagnetic latitude regions of low Earth orbit by observations with eyes oriented in the zenith and nadir directions. Due to the Earth's shadow and result that most light flashes are from heavy elements, there should be a measurable difference in observed light flashes when an astronauts look to nadir and zenith direction. The cosmic rays are scattered by the ISS walls, but the directions stay mostly similar. Thus, cosmic rays come more from the zenith than from the nadir direction. When astronauts look to zenith and to nadir, approximately the same amount of cosmic rays cross the eye, but when astronauts look to zenith, a larger retinal area is exposed. Roughly in zenith look it is 2/3 of the eye surface, in the nadir the area is smaller because part of the light flash photons escape through the cornea. Thus, there should be a different number of observed ligh flashes in nadir and zenith observation outside of SAA, if the mechanism behind light flashes in interplanetary and orbital regions excluding SAA is Cherenkov light production. 

\subsubsection{Shielding effect on Cherenkov light production}

In the previous sections, we showed that interactions of heavy cosmic ray nuclei with the human eye can lead to the visual perception of light flashes. In reality, light flashes were observed inside spacecraft or space stations, behind their walls, in pressurized internal space. Cosmic rays on their way to the human eye cross these walls. In this chapter, we investigate how this shielding affects the perception of light flashes. 

We simulated the interaction of cosmic rays with a human eye water model in the setup SG-ISS, with space station shielding representing the multilayer shielding of the International Space Station (ISS) walls with layers of aluminium and kevlar. The presented simulations should be regarded as an illustrative assessment of the influence of realistic ISS shielding rather than an exhaustive parametric study. The 50 GeV iron nucleus was selected as a representative case because iron nuclei produce the most intense Cherenkov events predicted by our model. Moreover, the majority of predicted iron induced light flashes originate from nuclei with kinetic energies of 50 GeV or higher. Integrating the AMS-02 iron spectrum shows that, of the predicted 0.17 iron nuclei crossing the eye per minute, approximately 0.14 nuclei per minute have energies exceeding 50 GeV.
Previous simulations demonstrated that the atmosphere around the water eye model in a box with a size similar to the size of ISS modules has a negligible effect on relativistic cosmic rays.

\begin{figure}[!htbp]
\noindent\includegraphics[width=\textwidth]{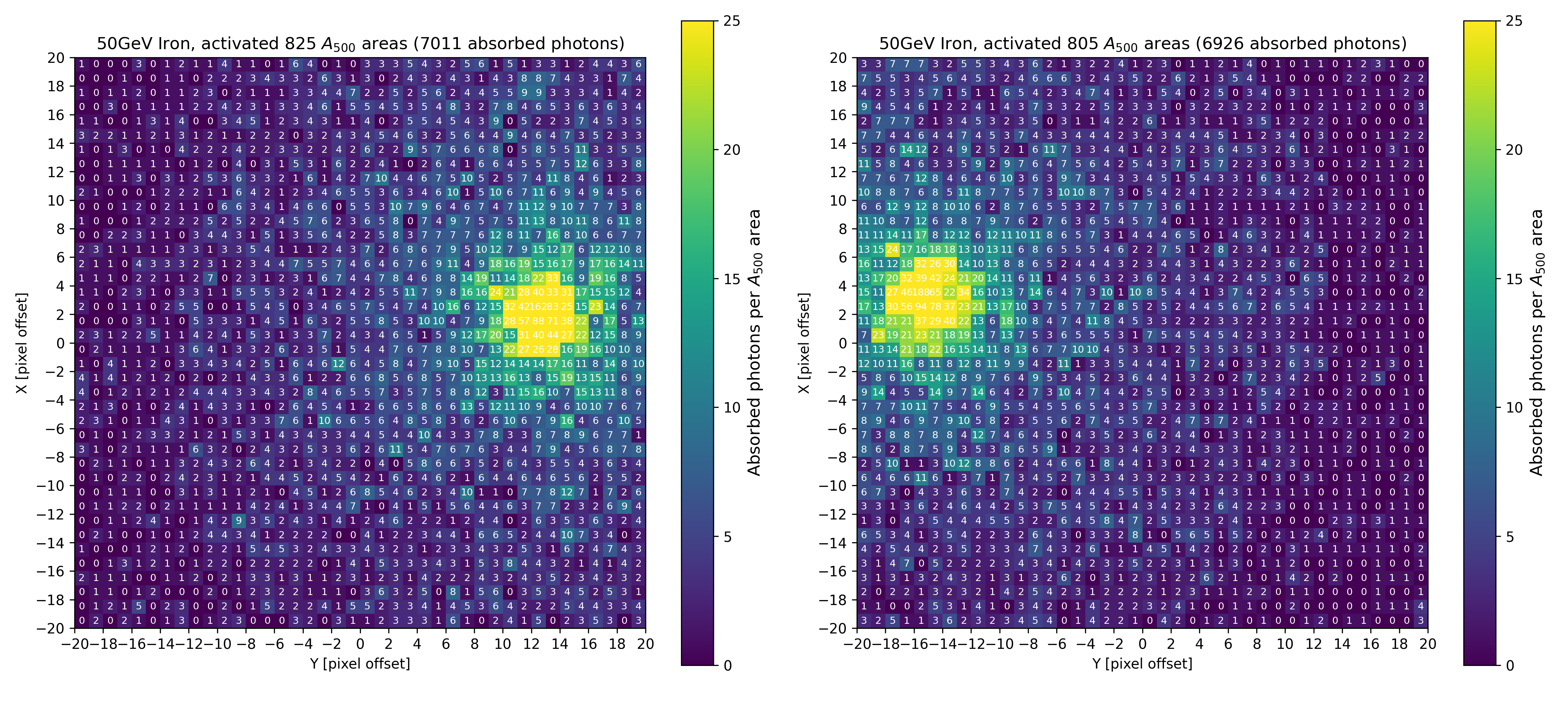}
\caption{Examples of retinal activation patterns produced by 50 GeV iron nuclei in Geant4 simulation setup SG-ISS with aluminium and kevral shielding. Colors indicate the number of absorbed photons in each $A_{500}$ area.}
\label{fig:50GeV_Iron_AlKevral_shielding_examples}
\end{figure}

The SG-ISS simulations show that the relativistic heavy nuclei producing Cherenkov light mostly cross the ISS walls with only a small reduction in energy. The primary 50 GeV iron nucleus loses on average 2.7 GeV (approximately 5\% of its initial kinetic energy) while traversing the aluminium and Kevlar shielding. An additional approximately 0.2 GeV is lost while traversing the air inside the ISS module. 
The simulated trajectories of the iron nuclei were slightly scattered, but most of them remained close to the central injection line. As two examples in figure \ref{fig:50GeV_Iron_AlKevral_shielding_examples} show, produced Cherenkov photons still activate more than 800 $A_{500}$ areas with approximately 7000 absorbed photons for the activation threshold $Ap \ge 5$. Because the shielding reduces the kinetic energy by only about 5\% in this representative example and introduces only modest angular scattering, the resulting Cherenkov photon distribution and retinal activation remain largely unchanged.

\begin{figure}[!htbp]
\noindent\includegraphics[width=\textwidth]{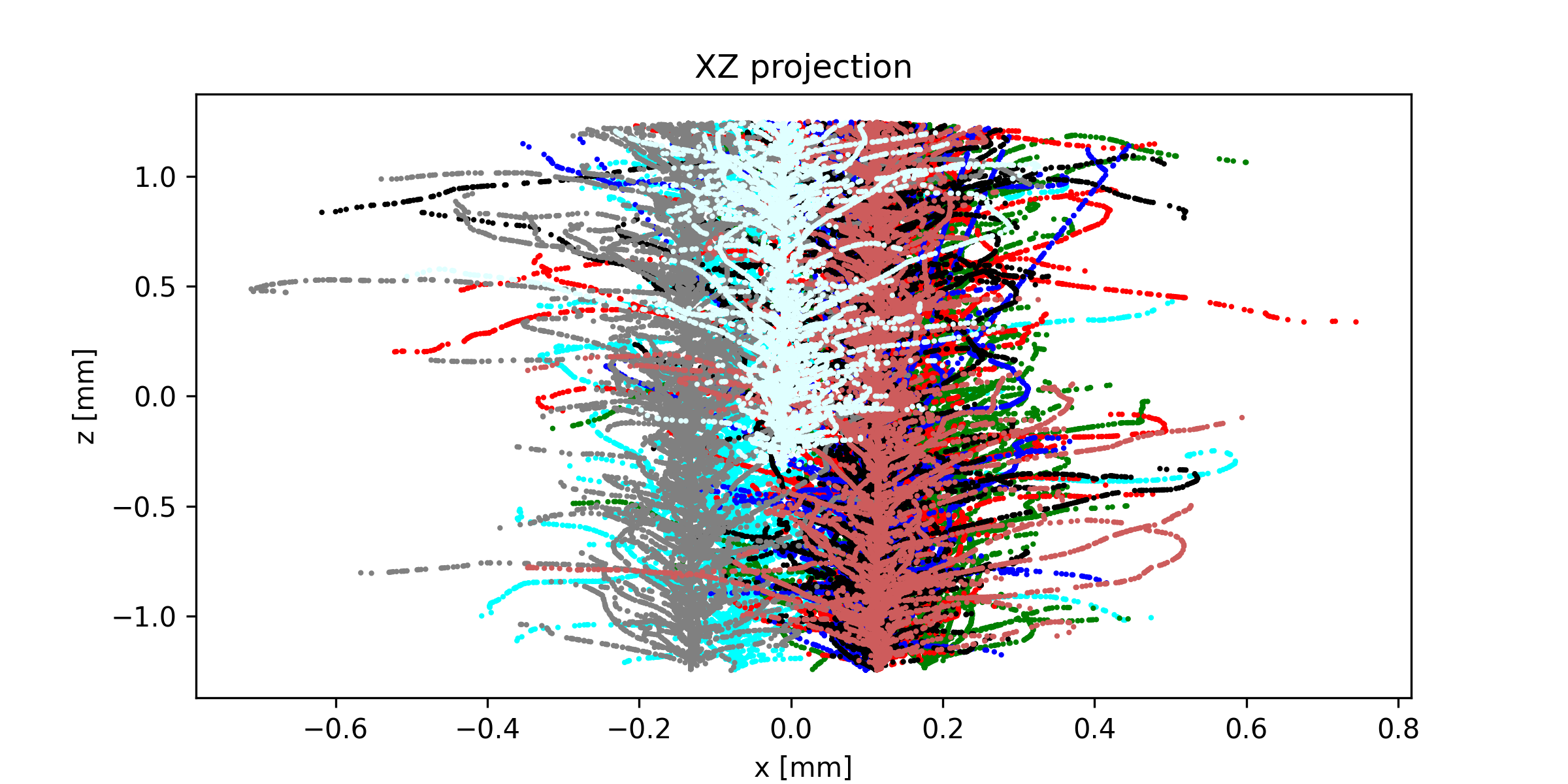}
\caption{Examples of visible Cherenkov photon production points for several trajectories of 50 GeV primary iron nuclei simulated with the SG-ISS configuration. Colors indicate different primary particles.}
\label{fig:XZ_projection_50GeV_iron_AlKevral_shielding}
\end{figure}

Figure \ref{fig:XZ_projection_50GeV_iron_AlKevral_shielding} shows the production points of Cherenkov photons along several representative trajectories of 50 GeV iron nuclei. Although the shielding slightly broadens the trajectories of the primary nuclei, the resulting angular deviations remain small. Since galactic cosmic rays arrive approximately isotropically over the exposed hemisphere in interplanetary space and remain nearly isotropic in low cutoff regions of low Earth orbit, this additional scattering does not significantly alter the flux of particles traversing the eye.

\subsubsection{Influence of surrounding head tissues}

The simplified eye model used throughout this work neglects the surrounding anatomical structures. To estimate the importance of this approximation, an additional Geant4 simulation was performed using a simplified head geometry consisting of concentric bone and brain spheres with an embedded water eye (setup SG-H\_HT in \ref{tab.2}). Point source beams were directed along the head symmetry axis toward the eye.

\begin{figure}[!htbp]
\noindent\includegraphics[width=\textwidth]{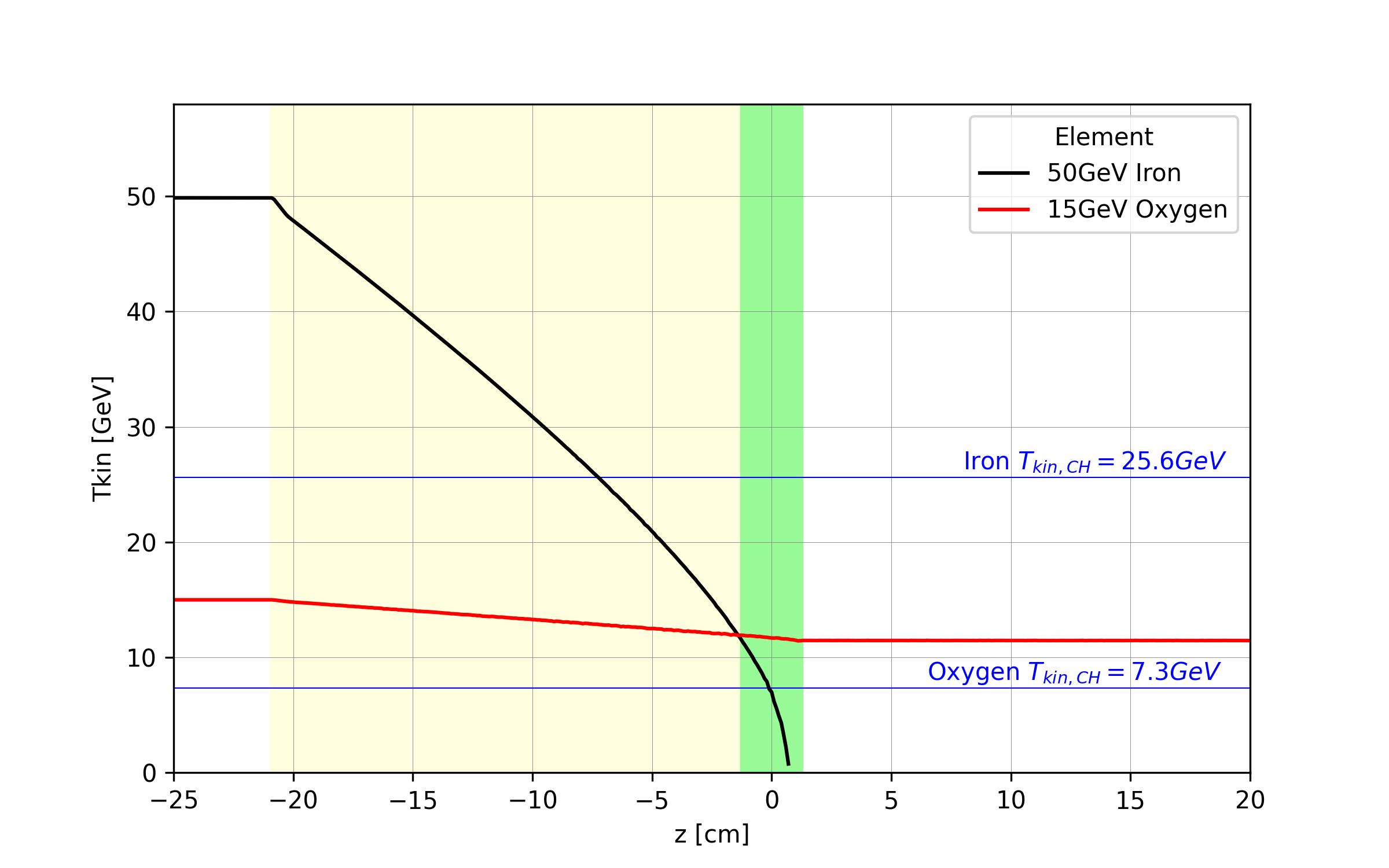}
\caption{volution of the average kinetic energy of 50 GeV iron and 15 GeV oxygen nuclei propagating through a simplified Geant4 head model consisting of bone, brain tissue, and a water filled eye. The initial energies were selected to be approximately twice the respective Cherenkov thresholds in water (horizontal blue lines). The shaded regions indicate the brain (yellow) and the eye (green). The 50 GeV iron nucleus loses energy rapidly, falls below the Cherenkov threshold before reaching the eye, and stops within the ocular volume, whereas the 15 GeV oxygen nucleus retains sufficient kinetic energy to traverse the eye while remaining above the Cherenkov threshold.}
\label{fig:50GeV_iron_15GeV_oxygen_Geant4_head_and_eye}
\end{figure}

Figure \ref{fig:50GeV_iron_15GeV_oxygen_Geant4_head_and_eye} shows the evolution of the average kinetic energy of 50 GeV iron and 15 GeV oxygen nuclei, selected to have initial energies approximately twice their respective Cherenkov thresholds in water (see table \ref{tab simple model}). The iron nucleus loses kinetic energy rapidly while traversing the head, falling below the Cherenkov threshold before entering the eye and stopping within the ocular volume. In contrast, the oxygen nucleus undergoes considerably smaller energy loss and enters the eye with kinetic energy still exceeding the Cherenkov threshold, allowing Cherenkov photons to be produced.

These results indicate that surrounding head tissues may substantially suppress Cherenkov light production for high charge heavy ions approaching the eye through long tissue paths, whereas lighter nuclei remain capable of producing Cherenkov radiation under similar conditions.

\subsubsection{Muon induced light flashes at the Earth's surface}
\label{Muon induced light flashes at the Earth's surface}

Using the methods and results from the previous analysis, we investigate why light flashes (LFs), commonly reported in space environments, are not observed at the Earth's surface. Since muons are the dominant component of secondary cosmic rays (CRs) at ground level, we repeated the Geant4 simulations using the SG-2.5 setup with a positive muon as the primary particle.

The mean number of photons produced in the water eye model by a 1 GeV primary positive muon was 640. This value is approximately 64\% higher than in the case of a 1 GeV proton. For a 100 MeV primary muon, the mean number of produced photons was 323. Positive muons exceed the speed of light in water for kinetic energies above approximately 52 MeV, which is approximately nine times lower than the corresponding threshold energy for protons.

Compared with protons in low Earth orbit, the intensity of muons at the Earth's surface is lower (approximately by a factor of 40), but muons are still sufficiently abundant and produce hundreds of visible photons while passing through the eye. The question therefore remains, why are light flashes not observed at the Earth's surface?

To evaluate the number of perceived light flashes at ground level, we used the positive muon spectrum measured by the BESS-97 experiment \cite{Rb13} at Lynn Lake (56.5$^{\circ}$N, 101.0$^{\circ}$W). This spectrum was selected because it covers a wide energy range, and the Lynn Lake location has low altitude and a low geomagnetic cut-off rigidity (0.4 GV).

When the isotropic intensity of positive muons was used, the resulting intensity of muons crossing the eye was approximately 19 per minute. Ground-level muons arrive only from the upper hemisphere and exhibit an angular distribution approximately proportional to $cos^2 \theta$. Accounting for both effects reduces the effective eye crossing rate to approximately one quarter of the isotropic full sphere estimate, resulting in approximately 5 muons crossing the eye per minute.

The Lynn Lake spectrum covers kinetic energies from 0.48 GeV to 20.45 GeV. The higher-energy part up to 500 GeV was extrapolated using a fit to the BESS 97 spectrum in the energy interval from 10 to 20 GeV. To obtain an upper limit estimate of the LF occurrence rate, the muon intensity between 52 MeV and 0.48 GeV was conservatively assumed to be constant and equal to the first measured BESS 97 energy bin. Under this assumption, the resulting intensity was approximately 24 muons crossing the eye per minute for isotropically incoming muons and approximately 6 per minute for a realistic ground level muon flux.

Assuming the most favorable observation geometry, with the observer looking in the zenith direction, the SG-2.5 configuration was oriented such that the z-axis of injected muons pointed toward the zenith. Using this configuration, we simulated 10,000 muons in Geant4 for kinetic energies of 70 and 80 MeV, and 0.1, 1, 10, 30, and 50 GeV.

Analysis of the Geant4 results showed that despite the production of hundreds of photons by each muon, no event resulted in activation of an $A_{500}$ retinal area, even for the lower threshold condition $Ap \ge 5$. In rare cases (2-4 events out of 10,000 simulated events at individual energies), an $A_{500}$ area accumulated 4 absorbed photons (assuming a photon absorption quantum efficiency of 29\%, consistent with all previous analyses). Since no event reached the activation threshold, an empirical upper limit of $P_i(T_{kin}, Ap) < 10^{-4}$ was adopted for the probability that an incident muon activates an $A_{500}$ retinal area by producing at least 5 absorbed photons.

When the muon probability function $P_i(T_{kin}, Ap)$ was included in the model using this upper limit probability for all energies, the predicted intensity of perceived light flashes was found to be below a few flashes per 10,000 minutes of observation. This corresponds to an upper limit estimate of approximately one muon induced light flash per 24 hours of continuous observation. The actual occurrence rate is likely substantially lower due to the activation threshold requirement. Moreover, any observed event would occur near the threshold of visual sensitivity and would most likely involve only a single activated $A_{500}$ area with approximately 5 absorbed photons.

The predicted occurrence rate is therefore more than four orders of magnitude lower than the light flash rates reported by astronauts, indicating that muon induced Cherenkov flashes are effectively unobservable at ground level.

At higher altitudes, such as high altitude observatories (2-4 km), muon intensities increase by only a factor of a few, which does not significantly alter the observability of muon induced light flashes. Even under these conditions, detection of muon induced Cherenkov flashes remains highly unlikely.

At even higher atmospheric altitudes, such as commercial aircraft flight altitudes (10-15 km), light flashes have been reported \cite{Rb7}. Although the muon intensity at these altitudes increases to values approaching an order of magnitude higher than at ground level, muon induced light flashes remain highly improbable. Therefore, light flashes observed at aircraft altitudes are more likely caused by heavy nuclei, similarly to observations made in space.







\section{Conclusion}

We developed a Cherenkov light production model that combines Geant4 simulations of cosmic ray interactions in the eye with retinal photon absorption, visual perception thresholds, and cosmic ray spectra to investigate the origin of astronaut light flashes. The model predicts that heavier cosmic ray nuclei, particularly iron, and to a lesser extent carbon and oxygen, produce sufficient Cherenkov photons to satisfy the modeled visual perception criteria, whereas hydrogen and helium generally do not. Geant4 simulations show that the number of produced Cherenkov photons ranges from a few hundred for hydrogen nuclei to tens of thousands for heavier nuclei across the investigated energy range. The simulated spatial distributions of Cherenkov photons were combined with retinal rod quantum efficiency, visual perception thresholds, and cosmic ray spectra to estimate the light flash rate associated with each cosmic ray element. 
The model predicts that light flashes observed by astronauts in interplanetary space and in low Earth orbit, except the South Atlantic Anomaly, are produced primarily by heavy nuclei, particularly iron, with additional contributions from carbon and oxygen. However, in the South Atlantic Anomaly, the Cherenkov light production model predicts a negligible light flash rate from trapped particles, indicating that the higher observed light flash rate in the South Atlantic Anomaly is due to another mechanism.

We also investigated the interaction of muons, the dominant secondary cosmic ray particles at Earth's surface, with the human eye. 
Although GeV muons produce approximately one third more visible Cherenkov photons than equally energetic protons, the spatial distribution of those photons on the retina is insufficient to satisfy the modeled perception criteria. This explains why light flashes are not observed at Earth's surface despite the high muon flux.

\section*{Data Availability}

The simulation model \textit{OpticApro-G4-10\_7} used in this study is available on GitHub \cite{ORs1}. This version was developed using Geant4 version 10.7 and is released under the BSD 2-Clause "Simplified" License. It may be freely used, modified, and distributed under the terms of this license. For more information, see the LICENSE file in the repository \cite{ORs1}. This version is also archived on Zenodo \cite{ORs2}. 

The code for model SG-Vac is also available on GitHub \cite{ORs3}, and archived on Zenodo \cite{ORs4}. This code is also under the BSD 2-Clause "Simplified" License.

Additionally, a newer version of the code, \textit{OpticApro-G4-11\_1} (developed for Geant4 version 11.1, but not used in this study), is publicly available on GitHub \cite{ORs5} and archived on Zenodo \cite{ORs6}.

The Cherenkov light production model is available on GitHub \cite{ORs7}, released under an MIT license and archived on Zenodo \cite{ORs8}.

No access restrictions apply to either version.

\section*{Acknowledgments}

The authors thank Jonathan N. Tinsley for valuable discussions on the minimum amount of light detectable by the human eye. We acknowledge the VEGA project 2/0124/25 for support.

\bibliography{references}     

\end{document}